\documentclass[twocolumn, 10pt]{IEEEtran}
\usepackage{hyperref} 
\usepackage[pdftex]{graphicx}
\usepackage{amssymb}
\usepackage{amsmath}
\usepackage{mathtools}
\usepackage{cite}
\usepackage{stfloats}
\usepackage{multirow}
\usepackage{subfigure}
\usepackage{psfrag}
\usepackage[mathscr]{euscript}
\usepackage{acronym}  
\usepackage{booktabs} 
\usepackage[table]{xcolor}
\usepackage{color}
\usepackage[utf8]{inputenc}
\usepackage{comment}
\usepackage{arydshln}
\usepackage{cleveref}

\acrodef{CCDF}{complementary cumulative distribution function}
\acrodef{CF}{characteristic function}
\acrodef{PPP}{Poisson point process}
\acrodef{CSI}{channel state information}
\acrodef{OFDM}{orthogonal frequency division multiplexing}
\acrodef{OFDMA}{orthogonal frequency division multiple access}

\acrodef{RV}{random variable}
\acrodef{i.i.d.}{independent, identically distributed}
\acrodef{PMF}{probability mass function}
\acrodef{PDF}{probability distribution function}
\acrodef{CDF}{cumulative distribution function}
\acrodef{ch.f.}{characteristic function}
\acrodef{AWGN}{additive white Gaussian noise}
\acrodef{SNR}{signal-to-noise ratio}
\acrodef{LRT}{likelihood ratio test}
\acrodef{DRT}{distance ratio test}
\acrodef{GLRT}{generalized likelihood ratio test}
\acrodef{CRLB}{Cram\'{e}r-Rao lower bound}
\acrodef{CRB}{Cram\'{e}r-Rao bound}
\acrodef{ZZLB}{Ziv-Zakai lower bound}
\acrodef{ZZB}{Ziv-Zakai bound}
\acrodef{LOS}{line-of-sight}
\acrodef{ToF}{time-of-flight}
\acrodef{NLOS}{non-line-of-sight}
\acrodef{GDOP}{geometric dilution of precision}
\acrodef{GPS}{Global Positioning System}
\acrodef{FIM}{Fisher information matrix}
\acrodef{PEB}{position error bound}
\acrodef{SPEB}{squared position error bound}
\acrodef{TOA}{time-of-arrival}
\acrodef{TOF}{time-of-flight}
\acrodef{WSN}{wireless sensor network}
\acrodef{MAC}{medium access control}
\acrodef{RSS}{received signal strength}
\acrodef{WAF}{wall attenuation factor}
\acrodef{TDOA}{time difference-of-arrival}
\acrodef{RF}{radiofrequency}
\acrodef{RTT}{round-trip time}
\acrodef{AOA}{angle-of-arrival}
\acrodef{MF}{matched filter}
\acrodef{ED}{energy detector}
\acrodef{ML}{maximum likelihood}
\acrodef{MSE}{mean-square error}
\acrodef{RMSE}{root-mean-square error}
\acrodef{LEO}{localization error outage}
\acrodef{ppm}{part-per-million}
\acrodef{ACK}{acknowledge}
\acrodef{UWB}{Ultrawide bandwidth}
\acrodef{TNR}{threshold-to-noise ratio}
\acrodef{LS}{least squares}
\acrodef{IR-UWB}{impulse radio UWB}
\acrodef{FCC}{Federal Communications Commission}
\acrodef{TH}{time-hopping}
\acrodef{PPM}{pulse position modulation}
\acrodef{MUI}{multi-user interference}
\acrodef{PDP}{power delay profile}
\acrodef{BPZF}{band-pass zonal filter}
\acrodef{SIR}{signal-to-interference ratio}
\acrodef{RFID}{radio frequency identification}
\acrodef{WPAN}{wireless personal area network}
\acrodef{WWB}{Weiss-Weinstein bound}
\acrodef{DP}{direct path}
\acrodef{MF}{matched filter}
\acrodef{MMSE}{minimum-mean-square-error}
\acrodef{SBS}{serial backward search}
\acrodef{SBSMC}{serial backward search for multiple clusters}
\acrodef{NBI}{narrowband interference}
\acrodef{WBI}{wideband interference}
\acrodef{INR}{interference-to-noise ratio}
\acrodef{CR}{channel response}
\acrodef{CIR}{channel impulse response}
\acrodef{CR}{channel  response}
\acrodef{RADAR}{radar}
\acrodef{MUR}{Multistatic radar}
\acrodef{JBSF}{jump back and search forward}
\acrodef{HDSA}{high-definition situation-aware}
\acrodef{RRC}{root raised cosine}
\acrodef{ST}{simple thresholding}
\acrodef{BTB}{Bellini-Tartara bound}
\acrodef{P-Max}{$P$-Max}  
\acrodef{MIMO}{multiple-input multiple-output}
\acrodef{MAP}{maximum a posteriori}
\acrodef{FG}{factor graph}
\acrodef{OP}{outage probability}
\acrodef{WED}{wall extra delay}
\acrodef{RMS}{root mean square}
\acrodef{SPAWN}{sum-product algorithm over a wireless network}
\acrodef{MDD}{minimum distance distribution}
\acrodef{MAP}{maximum a posteriori probability}
\acrodef{PAR}{probabilistic association rule}

\acrodef{ECU}{electronic control unit}
\acrodef{OBU}{on board unit}
\acrodef{CAN}{controller area network}
\acrodef{TPM}{trusted platform module}
\acrodef{HSM}{hardware security module}
\acrodef{KPS}{key predistribution system}
\acrodef{MAC}{message authentication code}
\acrodef{HMAC}{hash-based message authentication code}
\acrodef{CAN-FD}{CAN with flexible data rate}
\acrodef{V2V}{vehicle-to-vehicle}
\acrodef{ABE}{attribute based encryption}
\acrodef{IND-CPA}{indistinguishability under chosen plaintext attack}
\acrodef{SA}{security agent}
\acrodef{PRF}{pseudorandom function}
\acrodef{PRP}{pseudorandom permutation}
\acrodef{TA}{trust authority}
\acrodef{PPT}{probabilistic polynomial time}
\acrodef{ID}{identity}
\acrodef{PEAPOD}{privacy-enhanced attribute-based publishing of data}
\acrodef{EABEHP}{enhanced attribute-based encryption with hidden policy and credential}
\acrodef{MCU}{Micro Controller Unit}
\acrodef{ICS}{industrial control system}
\acrodef{C-IND-CPA-RUCA}{ciphertext indistinguishability against chosen plaintext attack and restricted user coalition attack}
\acrodef{P-IND-CPA-UCA}{policy indistinguishability against chosen plaintext attack and user coalition attack}

\usepackage{color}
\usepackage{dsfont}
\usepackage{bbm}

\newcommand{\Ws}[2]{{W_{}^{}}} 

\newcommand{\TSIR}[2]{{\tau_{}^{}}}

\DeclareMathAlphabet{\mathsf}{OML}{cmbr}{m}{it}

\newtheorem{definition}{Definition}
\newtheorem{theorem}{Theorem}

\newcommand{\bd}{\begin{description}}
\newcommand{\ed}{\end{description}}
\newcommand{\be}{\begin{enumerate}}
\newcommand{\ee}{\end{enumerate}}
\newcommand{\bi}{\begin{itemize}}
\newcommand{\ei}{\end{itemize}}
\newcommand{\bl}{\begin{list}}
\newcommand{\el}{\end{list}}
\newcommand{\bt}{\begin{tabbing}}
\newcommand{\et}{\end{tabbing}}

\newcolumntype{C}[1]{>{\centering\arraybackslash}p{#1}}

\begin{document}
	\newcommand{\paperTitle}{EC-SVC: Secure CAN Bus In-Vehicle Communications with Fine-grained Access Control Based on Edge Computing}
	
	\title{\paperTitle}
	
	\author{
		Donghyun Yu, Ruei-Hau Hsu, \textit{Member, IEEE}, and Jemin Lee, \textit{Member, IEEE} \vspace{-7mm}
	\thanks{
	Corresponding author is J. Lee.
	
	D.\ Yu and J. Lee are with
	Daegu Gyeongbuk Institute of Science and Technology (DGIST),
	333, Techno Jungang-daero, Daegu, Republic of Korea 42988
	(e-mail: \texttt{\{xaos4715, jmnlee\}@dgist.ac.kr}).
	
	R.\ H.\ Hsu is with National Sun Yat-sen University, 70
	Lienhai Rd., Kaohsiung 80424, Taiwan, R.O.C.
	(e-mail: \texttt{rhhsu@mail.cse.nsysu.edu.tw}).
	}
}

	\maketitle 
	
	\setcounter{page}{1}
	\acresetall
	
	\begin{abstract}	
		In-vehicle communications are not designed for message exchange between the vehicles and outside systems originally. Thus, the security design of message protection is insufficient. Moreover, the internal devices do not have enough resources to process the additional security operations. Nonetheless, due to the characteristic of the in-vehicle network in which messages are broadcast, secure message transmission to specific receivers must be ensured. With consideration of the facts aforementioned, this work addresses resource problems by offloading secure operations to high-performance devices, and uses attribute-based access control to ensure the confidentiality of messages from attackers and unauthorized users. In addition, we reconfigure existing access control based cryptography to address new vulnerabilities arising from the use of edge computing and attribute-based access control. Thus, this paper proposes an edge computing-based security protocol with fine-grained attribute-based encryption using a hash function, symmetric-based cryptography, and reconfigured cryptographic scheme. In addition, this work formally proves the reconfigured cryptographic scheme and security protocol, and evaluates the feasibility of the proposed security protocol in various aspects using the CANoe software.
	\end{abstract} 
	
	\vspace{-2.5mm}
	\section{Introduction}
	With the noticeable improvements in vehicles, internal devices in a vehicle start to share an amount of essential data of the car with each other.
	The most significant advantage of data sharing is to improve their data processing performance. For instance, electronic control unit (ECU), which is the most common machine sharing data in a car, can exchange information in order to make an important decision rapidly [1]. 
	Therefore, it is definitely indispensable for ECU to increase the amount of shared data for its decision performance in the car. Unfortunately, however, the increase in data to be shared among the internal devices does not always guarantee good results due to the existence of latent attackers whose objectives are to eavesdrop or manipulate the data for misbehaved operations. The message eavesdropping and modification attacks can be even effective against the current vehicles because the typical in-vehicle communications protocol does not provide data confidentiality and message authentication, which are the most basic requirements for secure communications [2]-[7].
	In particular, even the most representative in-vehicle communications protocol, that is, controller area network (CAN) protocol, has already been considered to be unable to satisfy the significant security requirements. This is mainly due to its obsoleteness, while it has been widely utilized in-vehicle communications [2], [8]. Moreover, CAN is widely used in industrial control system (ICS) including electronic equipments for aviation and navigation, medical devices and equipments, industrial automation and mechanical control, as well as vehicles. All of these systems are close to real-time systems, which can be fatal if security issues occur. 
	Hence, additional security mechanism such as the fine-grained access control of message exchange among different entities is urgently desired in CAN. 
	Therefore, to solve the security issues of CAN, many researchers have proposed new security protocols for CAN.
	The details of previous work are described in the following Sec. I-A.
	
	\vspace{-2.5mm}
	\subsection{Related Work}\label{sec:relatedwork}
	Initial works recognize the problem that ECUs have insufficient resources, such as power and computing capability to perform cryptographic protocols for secure in-vehicle communications [9]-[11]. Pierre et al. [10] presented the challenges of achieving high-security requirements with the insufficient power resources in in-vehicle networks. 
	Hisashi et al. [9] proposed an attestation-based security architecture for in-vehicle communications using the trusted platform module (TPM) in all ECUs. They designated resource-rich ECUs as master ECUs, and used the key predistribution system (KPS) for the authentication of the software configuration and the authenticated and encrypted communications.
	Hendrik et al. [11] provided a security solution the hardware security module (HSM) at ECUs and a key master. In the above mentioned papers, they dealt directly or indirectly the problems of the limited computing power of ECUs by using additional hardwares such as TPM and HSM in ECUs. However, as mentioned earlier, mounting additional hardwares on every ECUs is costly, which is practically impossible.
	
	Some works have also been proposed for in-vehicle communications security without mounting additional hardware at ECUs [12]-[15].
	Herrewege et al. [12] proposed the backward-compatible broadcast authentication protocol in CAN. 
	Bogdan et al. [13] provided a security protocol based on symmetric primitives using key splitting and message authentication code (MAC) mixing. They utilized nodes with high-computing power for distributing keys and a mixed MAC to achieve inter-group authentication, not one-to-one authentication. 
	Ryo et al. [14] focused on the problem of detecting spoofing messages in CAN bus, and proposed the lightweight authentication method. 
	Chung-Wei et al. [15] provided a security mechanism to keep CAN bus utilization low.
	In the above mentioned papers, they did not ensure the confidentiality of the data, so they are vulnerable to eavesdropping attacks. 
	
	Some recent works have achieved data confidentiality in in-vehicle communications [16], [17].
	Samuel et al. [16] provided a security protocol for secure real-time data processing in CAN through data encryption and authentication technology. The same authors also proposed a security architecture to meet the CAN with flexible data rate (CAN-FD) standards [17]. 
	This work achieved confidentiality, authentication, and integrity with the limited computing power of ECUs using cryptographic techniques with relatively low processing time such as symmetric key cryptography and hash functions. This work also did not provide a solution to the security issues that could occur if the high-performance node responsible for key management is corrupted by an attacker.
	
	To increase the security level of the in-vehicle communications, a security protocol using public key-based cryptography was provided for CAN-FD and FlexRay in [18]. However, this protocol is extremely hard to be executed while driving due to the limited performance of ECUs. Therefore, it should be executed in authorized garages or manufacturers, which, is not good in terms of usability. In addition, all ECUs sharing the same secret key can have the key exposure problem, caused by ECUs corrupted by attackers. Therefore, depending on the data being shared, the key must be shared only with the ECUs that need the data.
	\vspace{-3.5mm}
	\subsection{Motivation and Contribution}\label{sec:motivation}
	
	Despite of the intensive efforts for securing vehicle communications, most of the prior works have some issues as follows.	\emph{First, the excessive computing power is required.}
	Complicated operations are introduced to satisfy the aforementioned security requirements [9]-[12], [18]. 
	Generally, ECUs are unable to process the complex operations since they have lower computing capabilities than that of a typical computer.
	To reduce the security computation burden, it has also been proposed to mount additional components at ECUs such as hardware security modules, but this leads to additional costs.
	\emph{Second, the proposed protocols are not secure enough.}
	Some studies tried to construct their protocols by considering the real-time constraint of in-vehicle communications. However, they ended up meeting only low-level basic security requirements using symmetric-based cryptographic schemes [9], [11], [16], [17], [19].
	Therefore, for secure in-vehicle communications, it is desirable to develop new techniques that give lower computation burden to ECUs, while providing a higher level of security, which is the main objective of this paper.
	Therefore, this paper proposes a novel secure in-vehicle communications protocol with fine-grained access control based on edge computing, so-called EC-SVC, by exploiting the concept of edge computing to distribute the computation burden at ECUs [20]. 
	
	The contribution of this paper is as follows.
	\begin{itemize}
		\item \textit{EC-SVC} is proposed for in-vehicle communications that support data confidentiality, authentication, integrity, fine-grained access control, and policy and credential privacy.  
		\item We exploit the concept of edge computing by introducing the security agent (SA), which handles cryptographic operations on behalf of low computing power ECUs. In addition, we propose the \textit{enhanced attribute-based encryption with hidden policy and credential}, which achieves attribute-based access control with data confidentiality and policy privacy even against the SA.
		\item We formally prove the security of the enhanced attribute-based encryption and the proposed protocol. Furthermore, by implementing the proposed protocol in CAN, we show the feasibility of the proposed protocol under the real-time requirements in in-vehicle communications.
		\item By focusing on CAN among the in-vehicle network, we have shown that our work is not limited to the in-vehicle, but is also practical and feasible in a wide range of ICSs.
	\end{itemize}
	
	The remainder of this paper is organized as follows. Sec. II describes the system model, security requirements, and security models. Sec. III introduces the system preliminary including the newly proposed cryptographic scheme. Sec. IV describes the key management model and the proposed security protocols. Sec. V provides security analysis on proposed cryptographic scheme and the proposed protocol. Sec. VI analyzes the performance of the proposed protocol. Finally, Sec. VII presents a conclusion about this work.
	\section{System and Security Models}\label{sec:system}
	In this section, we describe the in-vehicle network model including the attack model and the security requirements of the network. We then introduce the security model and some definitions, which are used to analyze the security of the proposed protocol.
	\vspace{-2.5mm}
	\subsection{System Model and Security Requirements}\label{sec:systemmodel}
	
	\subsubsection{In-vehicle Network Model}
	
	\begin{figure}[t!]
		\begin{center}   
			{ 
				\includegraphics[width=1.00\columnwidth]{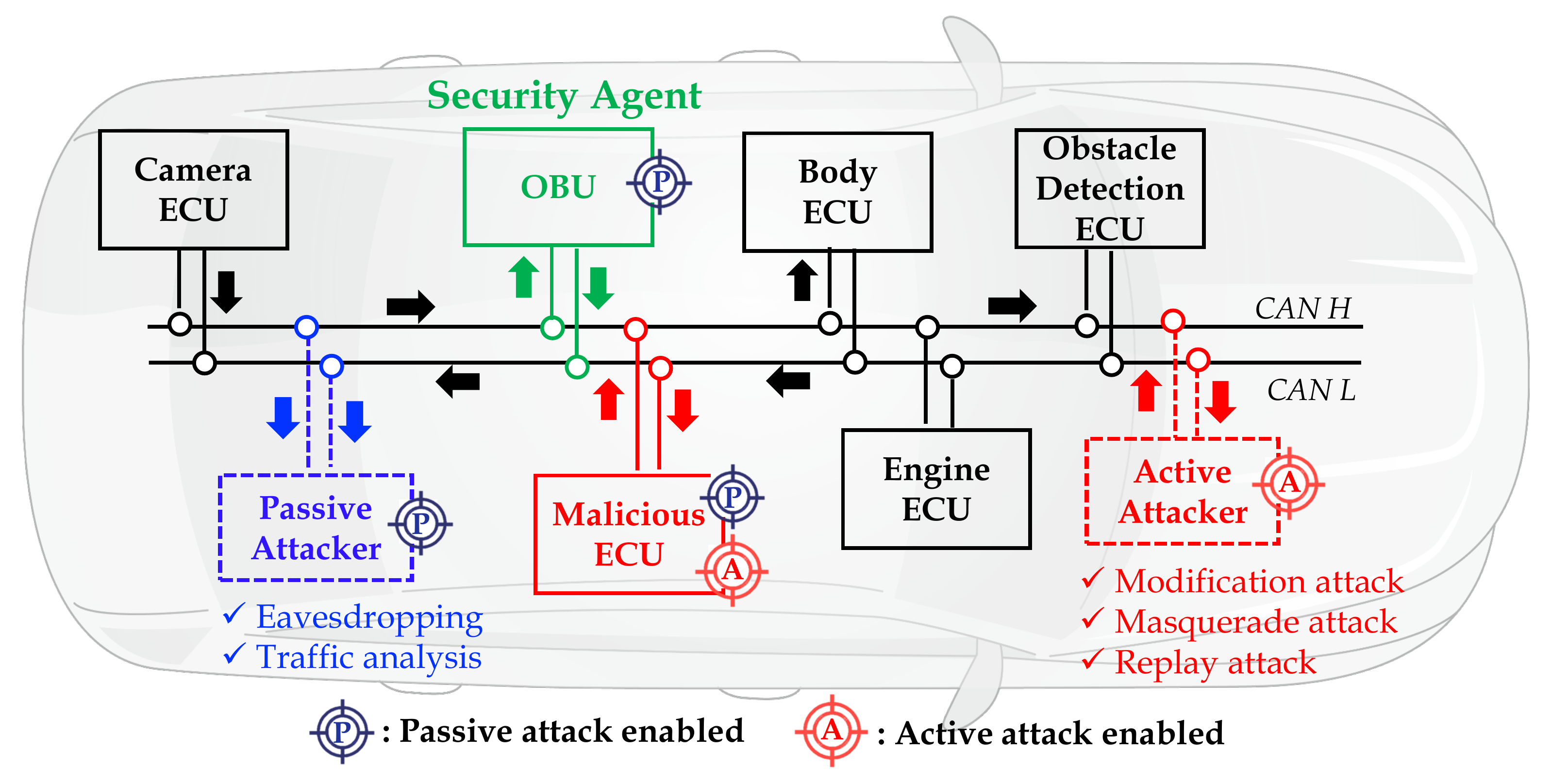}
			}
		\end{center}
	\vspace{-3.5mm}
		\caption{
			Overview of the system model.
		}
		\label{fig:SystemModel}
		\vspace{-4.5mm}
	\end{figure}
	We consider the in-vehicle network, where there exist two kinds of entities, ECU and on board unit (OBU) as shown in Fig. 1. All entities connect to CAN, which is a multi-master network, and communicate through the CAN bus. Each ECU collects the data from sensors, and shares it with actuators, other ECUs, and OBU. The ECU has low computing power, so long processing time is generally required to execute the advanced cryptography that provides a high-level of security. On the other hand, the OBU, which can communicate with nodes inside and outside of the vehicle, has higher computing power compared to the ECU. We utilize the OBU as an edge computing node, which allows to perform cryptography operations, offloaded by ECUs. We call this type of OBU as the SA [21], and assume the SA can be an honest-but-curious adversary. In this system model, we also consider security issues that may occur during offloading the cryptographic works from the ECU to the SA, and describe the details of the key management model for attribute-based key exchange in Sec. IV-A.
	\\
	\subsubsection{Security Requirements}
	
	As shown in Fig. 1, there can be two types of attackers in in-vehicle network, the passive attacker and the active attacker. The passive attackers can be connected to the CAN bus and eavesdrop the messages in CAN. The active attacker can cause severe problems to the vehicle by modifying the data sent by the ECU and the OBU or by sending a wrong key. In addition, the active attackers may corrupt the OBU and the ECU to obtain their secrets and use them to participate in communications for obtaining private data or session keys transmitted on the CAN. By considering the aforementioned attackers, we propose the following security requirements for the in-vehicle network and analyze these security requirements in Sec. V-B.
	\begin{itemize}
		\item \textit{Confidentiality} : 
		All messages are broadcasted due to the characteristics of the in-vehicle communications, so an attacker can easily eavesdrop the messages.	Hence, the confidentiality should be guaranteed, only the intended recipients can obtain the plaintext of the message.
		
		\item \textit{Authentication} : 	
		An attacker may retransmit the message, which exchanged between entities in the previous session, to impersonate a legitimate device. Therefore, we need to achieve mutual authentication to ensure that ECUs and SA can authenticate each other for the legitimacy.
		
		\item \textit{Fine-grained Access Control} : 
		Most of the transmitted data in a vehicle are not for all ECUs, but for certain ECUs.		
		In addition, when a node (e.g., ECU or OBU) is compromised by an attacker or some malicious nodes are connected in CAN, the transmitted message in CAN can be exposed without access control [22], [23]. Therefore, the fine-grained access control is required in in-vehicle networks.
		
		\item \textit{Policy and Credential privacy} : 
		A sender encrypts messages using the given policies to indicate the intended receivers according to their attributes for access control of message exchange. 
		The encryption and decryption procedures should not expose identity information of ECUs to avoid potential analysis or attacks to disrupt normal communications [22].
		
	\end{itemize}
\vspace{-2.5mm}
\subsection{Security Definitions of Enhanced Attribute-based Encryption with Hidden Policy and Credential}

To prove the security of the proposed attribute-based encryption scheme in Sec. III, so-called enhanced attribute-based encryption with hidden policy and credential (EABEHP), which achieves partial proxy decryption, hidden credential, and hidden policy against honest-but-curious decryption proxy, this section defines the security properties of EABEHP.

\begin{definition}[C-IND-CPA-RUCA] \label{def:1:C-IND-CPA-RUCA}
	EABEHP is said to be ciphertext indistinguishability against chosen plaintext attack and restricted user coalition attack (C-IND-CPA-RUCA) if any probabilistic polynomial time (PPT) adversary $\mathcal{A}$ has only negligible advantage to distinguish the ciphertext of two given messages in the following security game.
	\begin{enumerate}
		\item \textbf{Setup Phase:}		
		The challenger $\mathcal{C}$ sets up the EABEHP scheme and provides the attacker with all the public parameters of the system.
		\item \textbf{Training Phase 1:}		
		$\mathcal{A}$ can only corrupt either the proxy~(edge device) or any user except for the target user. That is, (i) the attacker can corrupt the proxy to learn its secrets and act on its behalf or (ii) $\mathcal{A}$ has the following abilities.
			\begin{itemize}
				\item $\mathcal{A}$ can register a new user or corrupt an honest user, who do not satisfy a policy $P^{*}$ on the system, thereby $\mathcal{A}$ can learn their secrets and act on their behalves.
				\item $\mathcal{A}$ can make requests of \textbf{TransformCipherText}, \textbf{Extract}, and \textbf{ProxyDecrypt1} to the proxy.
				\item $\mathcal{A}$ can make requests of \textbf{TimeKeyGen}, \textbf{TransformUserKey}, and \textbf{Shuffle} to honest users.
		\end{itemize}
		\item \textbf{Challenge Phase:} $\mathcal{A}$ outputs two messages $M_0$ and $M_1$ of equal length, and a policy $P^{*}$ under the following restriction. \\
		\textbf{Restriction 1:} None of the corrupted users in \textbf{Training Phase 1} satisfy $P^{*}$. \\
		$\mathcal{C}$ then flips the random coin $b \stackrel{R}{\in} \{0,1\}$ and generates $C^{*}$ by encrypting the message $M_b$ under the policy $P^{*}$ by the \textbf{Encrypt} algorithm according to $b$. $\mathcal{C}$ returns $C^{*}$ to $\mathcal{A}$.
		\item \textbf{Training Phase 2:}
		$\mathcal{A}$ can perform the operations defined in \textbf{Training Phase 1}, except that none of the corrupted users can satisfy the policy $P^*$.
		\item \textbf{Guessing Phase:}
		$\mathcal{A}$ outputs a guessing $b' \stackrel{R}{\in} \{0,1\}$. $\mathcal{A}$ wins the game if $b' = b$.
		\\
	\end{enumerate}
\end{definition}
\vspace{-2.5mm}
	\begin{definition}[P-IND-CPA-UCA]\label{def:2:P-IND-CPA-UCA}
		EABEHP is said to be policy indistinguishability against chosen plaintext attack and user coalition attack (P-IND-CPA-UCA) if all PPT adversaries only have negligible advantage to distinguish the ciphertext of two given policies in the following security game.
		\begin{enumerate}
			\item \textbf{Setup Phase:} Same as that in the Definition 1.
			\item \textbf{Training Phase 1:} Same as that in the Definition 1.
			\item \textbf{Challenge Phase:} $\mathcal{A}$ sends a chosen message $M^*$ and two chosen policies, $P_0$ and $P_1$, for the encryption of $M^*$ under the following restriction. \\
			\textbf{Restriction 2:} All the corrupted users satisfy none of the policies, $P_0$ and $P_1$, or they all satisfy both policies. 
			$\mathcal{C}$ selects a random bit $b\in\{0,1\}$ and encrypts $M^*$ with the given $P_b$ to generate $C^*$ according to $b$. 
			\item \textbf{Training Phase 2:} Same as that in the \textbf{Training Phase 1}.
			\item \textbf{Guessing Phase:} Same as that in the Definition 1.
		\end{enumerate}
\end{definition}

	\begin{definition}[Credential Privacy]\label{def:3:Credential-Privacy}
		EABEHP is said to support credential privacy if all PPT adversaries only have negligible advantage to distinguish the real credential of the specified user from the credentials of the other users with non-negligible advantages as the following game. 
		\begin{enumerate}
			\item \textbf{Setup Phase:} Same as that in the Definition 1.
			\item \textbf{Training Phase 1:} Same as that in the Definition 1.
			\item \textbf{Challenge Phase:} $\mathcal{A}$ outputs the credentials of two users, a selected policy $P^*$, and a selected message $M$. $\mathcal{C}$ then outputs a ciphertext of $M$ with $P^*$ and $SK_{U_b}$ according to $b\stackrel{R}{\in}\{0,1\}$, where the associated attributes of $SK_{U_1}$ and $SK_{U_2}$ either both satisfy $P^*$ or do not satisfy $P^*$.
			\item \textbf{Training Phase 2:} Same as that in the \textbf{Training Phase 1}, except that the corruption of the users possessing $SK_{U_1}$ or $SK_{U_2}$ is not allowed.
			\item \textbf{Guessing Phase:} Same as that in the Definition 1.
		\end{enumerate}
	\end{definition}
\vspace{-2.5mm}
\subsection{Security Definitions of Secure In-Vehicle Communications with Access Control}\label{sec:securitymodel}

We capture the capabilities of attackers by following definitions in the system model.
We first explain the notation used in the security model.
The proposed protocol is called $\Gamma$, and we regard the communication between two users $A$ and $B$ in communication sessions at $t_1$ and $t_2$ as $\Gamma_{A,B}^{t_1}$ and $\Gamma_{B,A}^{t_2}$, respectively. 
We describe oracles, propose attackers who can query these oracles, and define the security of the proposed protocols according to security requirements.

The oracles used to capture the attacker's capabilities are as follows.
\begin{itemize}
	\item \textbf{Execute}($\Gamma_{A,B}^{t_1}$, $\Gamma_{B,A}^{t_2}$) : This oracle models all kinds of passive attackers that can eavesdrop all data between $\Gamma_{A,B}^{t_1}$ and $\Gamma_{B,A}^{t_2}$.
	
	\item \textbf{Send}($\Gamma_{A,B}^{t_1}$, $M$) : This oracle models an active attacker that sends a message $M$ to $\Gamma_{A,B}^{t_1}$.
	
	\item \textbf{Expose}($\Gamma_{A,B}^{t_1}$) : This oracle models the exposure of the session key of $A$, shared with $B$, at communication session $t_1$.
	
	\item \textbf{Corrupt}($\Gamma_{A,B}^{t_1}$) : This oracle models the exposure of the long-term secret key of $A$, shared with $B$, at communication session $t_1$.
	
	\item \textbf{Test}($\Gamma_{A,B}^{t_1}$): This oracle models the test of session key security. When one queries this oracle and both parties, which are the partners of each other, in the protocol are accepted, it will return a real session key or a random string depending on a random bit. Otherwise, it returns an invalid output.
	
	\item \textbf{TestPolicy}($P_0, P_1, M$) This oracle models to test the privacy of the given policies for encryption. When one queries this oracle with the inputs of two given policies, $P_0$ and $P_1$, and a message $M$, it will output an encryption on $M$ with $P_b$ according to the randomly selected $b\in\{0,1\}$.
	
	\item \textbf{TestCert}($\Gamma_{A,B}^{t_1}, P^*$): This oracle models to test the privacy of user's credential. When one queries this oracle with the input of $\Gamma_{A,B}^{t_1}$ and the target policy $P^*$, it will output either the credential of $\Gamma_{A,B}^{t_1}$ or a randomly selected credential with the restriction that the attributes of both credentials satisfy $P^*$ or do not satisfy $P^*$.

\end{itemize}
We also define the security properties of the proposed EC-SVC according to the security requirements discussed in Sec. II-A as follows.
	
	\begin{definition}[Mutual Authentication] \label{def:4:Mutual-authentication}
	We assume that $S$ simulates $\Gamma_{A,B}^{t_1}$ and $\Gamma_{B,A}^{t_2}$, and interacts with $\mathcal{A}$, who can query polynomial number of \textbf{Execute} and \textbf{Send} oracles.
	After $\mathcal{A}$ queries these oracles, it sends a message to be accepted by $\Gamma_{A,B}^{t_1}$ or $\Gamma_{B,A}^{t_2}$, where $\Gamma_{A,B}^{t_1}$ or $\Gamma_{B,A}^{t_2}$ has not accepted each other. $\mathcal{A}$ has the following advantage
	\begin{align}
	\text{Adv}_{\mathcal{A}}^{\text{MuAuth}}=\text{Pr}[\mathcal{A}\ \text{accepted by}\ \Gamma_{A,B}^{t_1}\ \text{or}\ \Gamma_{B,A}^{t_2}].
	\end{align}
	The mutual authentication between $A$ and $B$ is guaranteed if $\text{Adv}_{\mathcal{A}}^{\text{MuAuth}}$ is negligible.
	\end{definition}
	
	\begin{definition}[Attribute-based Key Exchange] \label{def:5:Attribute-based-Key-exchange}
	There are $S$ simulating $\Gamma_{A,B}^{t_1}$ and $\Gamma_{B,A}^{t_2}$, and $\mathcal{A}$, who can query polynomial number of \textbf{Execute} and \textbf{Send} oracles in polynomial time.
	After $\mathcal{A}$ queries these oracles, if $\Gamma_{A,B}^{t_1}$ and $\Gamma_{B,A}^{t_2}$ are accepted by each other with a session key $K$, $\mathcal{A}$ queries \textbf{Test} to obtain a session key $K$ or a random string according to a random bit $b \in \{0,1\}$. $\mathcal{A}$ has the following advantages
	\begin{align}
	\text{Adv}_{\mathcal{A}}^{\text{AKE}}=\text{Pr}[\text{Succ}_{\mathcal{A}}^{\text{AKE}}]\ -\ 1/2,
	\end{align}
	where $\text{Succ}_{\mathcal{A}}^{\text{AKE}}$ is the event that $\mathcal{A}$ outputs a guess $b' = b$.
	If $\text{Adv}_{\mathcal{A}}^{\text{AKE}}$ is negligible, the attribute-based key exchange security is achieved.
	\end{definition}

	\begin{definition}[Policy Privacy] \label{def:6:Policy-Privacy}
	$S$ simulates $\Gamma_{A,B}^{t_1}$ and $\Gamma_{B,A}^{t_2}$, and interacts with $\mathcal{A}$, who can query polynomial number of \textbf{Execute} and \textbf{Send} oracles in polynomial time. After this phase, $\mathcal{A}$ queries \textbf{TestPolicy} with message $M$ and two valid policies, $P_0$ and $P_1$, as the input to obtain a $C_0$ or $C_1$ according to a random bit $b \in \{0,1\}$, where $C_0$ and $C_1$ are encryption for $M$ with $P_0$ and $P_1$, respectively. $\mathcal{A}$ has the following advantages
	\begin{align}
	\text{Adv}_{\mathcal{A}}^{\text{PP}}=\text{Pr}[\text{Succ}_{\mathcal{A}}^{\text{PP}}]\ -\ 1/2,
	\end{align}
	where $\text{Succ}_{\mathcal{A}}^{\text{PP}}$ is the event that $\mathcal{A}$ outputs a guess $b' = b$.
	If $\text{Adv}_{\mathcal{A}}^{\text{PP}}$ is negligible, the policy privacy is achieved.
	\end{definition}
	
	\begin{definition}[Credential Privacy] \label{def:7:Credential-Privacy}
	$S$ simulates $\Gamma_{A,B}^{t_1}$ and $\Gamma_{B,A}^{t_2}$, and interacts with $\mathcal{A}$, who can query polynomial number of \textbf{Execute} and \textbf{Send} at polynomial time. After this phase, $\mathcal{A}$ queries \textbf{TestCert} with target policy $P^*$ as the input to obtain a credential of $\Gamma_{A,B}^{t_1}$ or a randomly selected credential according to a random bit $b \in \{0,1\}$, where the attributes of both credentials satisfy $P^*$ or do not satisfy $P^*$. $\mathcal{A}$ has the following advantages
	\begin{align}
	\text{Adv}_{\mathcal{A}}^{\text{CP}}=\text{Pr}[\text{Succ}_{\mathcal{A}}^{\text{CP}}]\ -\ 1/2,
	\end{align}
	where $\text{Succ}_{\mathcal{A}}^{\text{CP}}$ is the event that $\mathcal{A}$ outputs a guess $b' = b$.
	If $\text{Adv}_{\mathcal{A}}^{\text{CP}}$ is negligible, the credential privacy is achieved.
	\end{definition}
\vspace{-2.5mm}
	\section{System Preliminaries}\label{sec:preliminaries}
	
	This section introduces the required background for the proposed protocol including the cryptographic algorithms used in the protocol.
	
	\subsection{Pseudorandom Function and Permutations}\label{sec:pseudorandom}
	We describe the hash function, e.g., SHA-256, and the symmetric encryption, e.g., AES128, used in this protocol as pseudorandom function (PRF) and pseudorandom permutation (PRP) [24], respectively. First, the PRF [25], [26] defined over $(K, X, Y)$ is an efficient and deterministic function, which returns a pseudorandom output sequence
	\begin{align}
	H : K_H \times X \rightarrow Y,
	\end{align}
	where $K_H$ is the key space, $X\subseteq \{0,1\}^{l_1}$ is the input space, $Y\subseteq \{0,1\}^{l_2}$, and $l_1 > l_2$.
	The PRP [27] defined over $(K, X)$ is an efficient and deterministic function which returns a pseudorandom output sequence
	\begin{align}
	E : K_E \times X \rightarrow X',
	\end{align}
	where $K_E$ is the key space, $X\subseteq \{0,1\}^{l}$ is the input space, and $X'\subseteq \{0,1\}^{l}$. There is an efficient inversion algorithm $D(K_E , X')$ for this PRP. We additionally describe the shuffle function, which is a kind of pseudorandom permutation to be used in the EABEHP scheme. The shuffle function has the same input and output space, which returns a pseudorandom output sequence
	\begin{align}
	SH : K_{SH} \times X \rightarrow X,
	\end{align}
	where $K_{SH}$ is the key space, and $X\subseteq \{0,1\}^{l_N}$ is the input and output space. There is an efficient inversion algorithm $SH^{-1}(K_{SH} , X)$ for this shuffle function.
	\vspace{-2.5mm}
	\subsection{Enhanced Attribute-based Encryption with Hidden Policy and Credential}\label{sec:EABEHP}
	
	\subsubsection{Intuition} The proposed EABEHP is an enhanced scheme of the privacy-enhanced attribute-based publishing of data (PEAPOD) scheme in [22]. In order to leverage edge computing resources, we use the concept of the security agent (SA), which is the device with more powerful computational resource and performs the cryptographic operations, offloaded by resource-limited devices [20]. Since we consider the SA as an honest-but-curious, the EABEHP needs to satisfy two additional security properties: 1) confidentiality against SAs for proxy decryption, and 2) hidden attributes and policies against SAs. 
	
	First, to achieve the confidentiality of encrypted message against the proxy, who performs \textbf{ProxyDecrypt1} with the given ciphertext, $sC'_{r}$ and user secret keys, ($AK_{U_r}$, $RK_{U_r}$), $sC'_{r}$ is shuffled by the shuffle function, \textbf{Shuffle}, which takes the time key $\omega_{k}$ of each time slot $k$. Since $\omega_{k}$ is unknown to the proxy, the proxy cannot recover the original ciphertext $C$ from the shuffled $sC'_{r}$ and decrypt the ciphertext successfully.
	
	Second, to conceal the policies of encryption from the proxy and outsiders, the policy is used to decide to encrypt a message tuple, which is the divided partial message, i.e., $k_i$, a random value $\alpha_i$, or 1, depending on that the $i$-th attribute is required, unrequired, or irrelevant specified in the policy. Thus, when any attacker wants to distinguish if a ciphertext is encrypted with which one of two given policies, where one of them will be randomly selected as the policy of the encryption, the attacker has to decrypt the encrypted message tuple first. Otherwise, all the encrypted message tuples of the ciphertext are considered as random variables. The attacker will not learn any policy information if the confidentiality of the message tuples based on the ElGamal encryption is guaranteed.
	
	 Third, since the proxy for partial decryption process, i.e., \textbf{ProxyDecrypt1}, needs to know the attribute indices of each receiver, it may expose user attribute information. Thus, a permuted attribute indices, $\hat{I}_r$, for the receiver $r$ will be given to the proxy for partial decryption. Since \textbf{ProxyDecrypt1} takes $\hat{I}_r$ for the partial decryption proceess, the output will remain $p_i \times{}g^{r_j (s_i-s_i')}$ for each message tuple, where $i$ is the inverse permuted attribute index and $i'$ is permuted $i$ using a $SH_{\omega_k}$. Thus, the receiver, who knows $\omega_k$, can calculate $\hat{I}_r$ and use the tuples, $AM_{r_j}=\{g^{r_j s_i}\}_{i\in{}I}$, for the receiver to cancel $g^{r_j (s_i-s_i')}$ to recover each message tuple $p_i$ during the decryption procedure by \textbf{ProxyDecrypt2}.
	 
	 The proposed scheme consists of ten algorithms.
	\begin{itemize}
		\item \textbf{Setup($1^{\lambda}$)}: This algorithm chooses the cyclic group $\mathbb{G}$ of prime order $p$ with a generator $g$. Next, it chooses a large prime number $q$ such that $q|(p-1)$ and random numbers $\{a_i\}_{i\in{}I}$ for all attributes of the system, where $I=\{1, 2, 3, \ldots , N\}$ is the universal set of attribute indices of the system, and $N$ is the number of system attributes. After that, the algorithm generates the master public key, $MPK$, of the system as
		\begin{align} \notag
		MPK=\{g, \; p, \; q, \; \{ PK_{S_i}=g^{a_i}\}_{i\in{}I}\},
		\end{align}
		where $a_i \in \mathbb{Z}^{*}_q$. Then, it randomly chooses a master secret key $MSK=K_S$ and generates a transformation secret key $TK=\{ TK_{S_i}=s_i\}_{i\in{}I}$ where $a_i+s_i=K_S$ for all $i\in I$. Finally, it generates a group key $K_{\text{group}}$ for all users in the system. The algorithm then outputs $MSK$, $MPK$, $TK$, and $K_{\text{group}}$.
		
		\item \textbf{KeyGen}($MSK,\text{ID}_j,I_j$): This algorithm generates user attribute keys, $SK_{U_j}=\{SK_{U_{j,i}}=a_{j,i}\}_{i\in I_j}$, where $I_j \hspace{-1mm} \subseteq \hspace{-1mm} I$ is the set of attributes indices of user $j$. Next, it generates a re-encryption key $RK_{U_j}=\sum_{i\in I_j} s_{j,i}$ for all attribute indices of the user $j$ where $a_{j,i}+s_{j,i}=K_S$. This algorithm outputs $SK_{U_j}$ and $RK_{U_j}$ for user $j$.
		
		\item \textbf{TimeKeyGen}($K_{\text{group}},t_{k}$): This algorithms takes $K_{\text{group}}$ and the time slot $t_{k}$ as inputs, and generates $\omega_{k}=H(r_{k}||K_{\text{group}})$ as the output, where $r_k$ is randomly selected and distinct for different $t_k$.
		
		\item \textbf{TransformUserKey($\omega_k,SK_{U_j}$)}: This algorithm takes $\omega_k$ and $SK_{U_j}$ as inputs, and outputs $AK_{U_j}=\sum_{i\in I_j}a_{j,i}+\omega_k$ as the transformed user attribute key.
		
		\item \textbf{Encrypt($MPK, T, \omega_k, M$)}: This algorithm first takes $MPK$, $T$, $\omega_{k}$, and $M$ as inputs, and outputs $C$ as the ciphertext of $M$. Here, $T=\{t_i\}_{i\in I}$ is a policy set, where $t_i = 1$ if the attribute $i$ is required, $t_i = 0$ if the attribute $i$ is irrelevant, and $t_i = -1$ if the attribute $i$ is unrequired, and $M\in \mathbb{Z}_q$ is the message to be encrypted. Then, it generates the message tuples depending on each $i\in{}I$ as
		\begin{align}
		\notag
		p_i=
		\begin{cases} k_i & \text{if} \; t_{i}=1
		\\ 1 & \text{if} \; t_{i}=0
		\\ \alpha_i & \text{if} \; t_{i}=-1
		\end{cases},
		\end{align} 
		\begin{align} \notag
		\text{such that} \prod_{\substack {t_i\in T \wedge t_i = 1, \\ \forall i\in I}} p_i\equiv M \; (mod \; q),
		\end{align}
		for randomly selected $\alpha_i\in {\mathbb{Z}}_q$. Next, it randomly selects $r_j\in \mathbb{Z}_q$ and encrypts each message tuple, $p_i$, with $PK_{S_i} = g^{a_i}$ as
		\begin{align} \notag
		C= \{A, {\langle B_i \rangle}_{i \in I}, D\}=\{g^{r_j}, {\langle p_i (g^{a_i})^{r_j}\rangle}_{i \in I}, (g^{r_j})^{\omega_k}\},
		\end{align}
		where $\langle \cdot \rangle$ means a sequence.
		
		\item \textbf{Shuffle($C , \omega_k$)}: This algorithm permutes the order of tuples by a pseudorandom permutation as
		\begin{align} \notag
		sC &= \{A, {\langle \widehat{B}_{i}\rangle}_{i \in I}, D\} \\ \notag
		&=\{g^{r_j}, {\langle \widehat{B}_{i} = p_{i'} (g^{a_{i'}})^{r_j}\rangle}_{i \in I \wedge i'=SH_{\omega_k}(i)}, (g^{r_j})^{\omega_k}\},
		\end{align}
		where $i'=SH_{\omega_k}(i)$, is a shuffle function, which takes $\omega_k$ and $i\in I$ as inputs and outputs $i' \in I$.
		
		\item \textbf{TransformCipherText($sC , TK$)}: This algorithm use $TK$ to transform $sC$ as
		\begin{align} \notag
		sC' &= \{A', {{\langle B'_{i}}\rangle}_{i \in I}, D'\} = \{A, {\langle \widehat{B}_{i} A^{s_i}\rangle}_{i \in I}, D\} \\ \notag
		&= \{g^{r_j}, {\langle p_{i'} g^{a_{i'}r_j+s_i r_j}\rangle}_{i \in I \wedge i'=SH_{\omega_k}(i)}, (g^{r_j})^{\omega_k}\}.
		\end{align}
		
		\item \textbf{Extract($sC' , \hat{I}_r$)}: This algorithm takes $sC'$ and $\hat{I}_r$ as inputs and outputs the extracted ciphertext $sC'_r$. Here, $\hat{I}_r$ is a set of elements obtained by inversely permuting each element $i \in I_r$ such as $SH^{-1}_{\omega_k}(i)$. It extracts $sC'_r$ from $sC'$ according to $\hat{I}_r$ as
		\begin{align} \notag 
		sC'_r &= \{A'_r, B'_r , D'_r \} = \{A', \prod_{\substack {i \in \hat{I}_r}} B'_{i}, D'\} \\ \notag
		&= \{A, \prod_{\substack {i \in \hat{I}_r}} \widehat{B}_{i}A^{s_{i}}, D\} \\ \notag
		&=\{g^{r_j}, \hspace{-3.5mm}\prod_{\substack {i \in \hat{I}_r \wedge \\ i'=SH_{\omega_k}(i)}} \hspace{-3.5mm}p_{i'} g^{a_{i'}r_j+s_i r_j}, (g^{r_j})^{\omega_k}\}.
		\end{align}
		
		\item \textbf{ProxyDecrypt1($sC'_r , AK_{U_r} , RK_{U_r}, TK$)}: This algorit- hm takes $sC'_r$, $AK_{U_r}$, and $RK_{U_r}$, and $TK$ as inputs, and outputs the partial decrypted ciphertext $sC''_r$ and decryption materials $AM_{r_j}$ as
		\begin{align} \notag
		sC''_r &=D'_r\cdot B'_r/(A'_r)^{AK_{U_r}+RK_{U_r}} \\ \notag
		&= \bigg(\hspace{-3.5mm}\prod_{\substack {i \in \hat{I}_r \wedge \\ i'=SH_{\omega_k}(i)}} \hspace{-3.5mm}p_{i'}\bigg)(g^{r_j})^{\sum_1 s_i-s_{i'}}, \\ \notag
		& AM_{r_j} = \{(A'_r)^{s_i}\}_{i \in I} = \{g^{r_j s_i}\}_{i \in I},
		\end{align}
		where $N_r$ is the number of attributes of user $r$, $\sum_1 = \sum_{i\in \hat{I}_r \wedge i'=SH_{\omega_k}(i)}$.
		
		\item \textbf{ProxyDecrypt2($sC''_r , AM_{r_j} , I_r$)}: This algorithm takes $sC''_r$, $AM_{r_j}$, and $I_r$ as inputs, and outputs $M$ as
		\begin{align} \notag
		M = sC''_r \cdot \bigg(\prod_{i \in I_r}g^{r_j s_i}\bigg) \bigg/ \bigg(\prod_{i \in \hat{I}_r} g^{r_j s_{i}}\bigg).
		\end{align}
 
	\end{itemize}

	\section{Proposed Secure In-Vehicle Communications with Fine-grained Access Control based on Edge Computing~(EC-SVC)}\label{sec:ASVC}

	\begin{figure}[t!]
	\begin{center}   
		{ 
			\includegraphics[width=1.00\columnwidth]{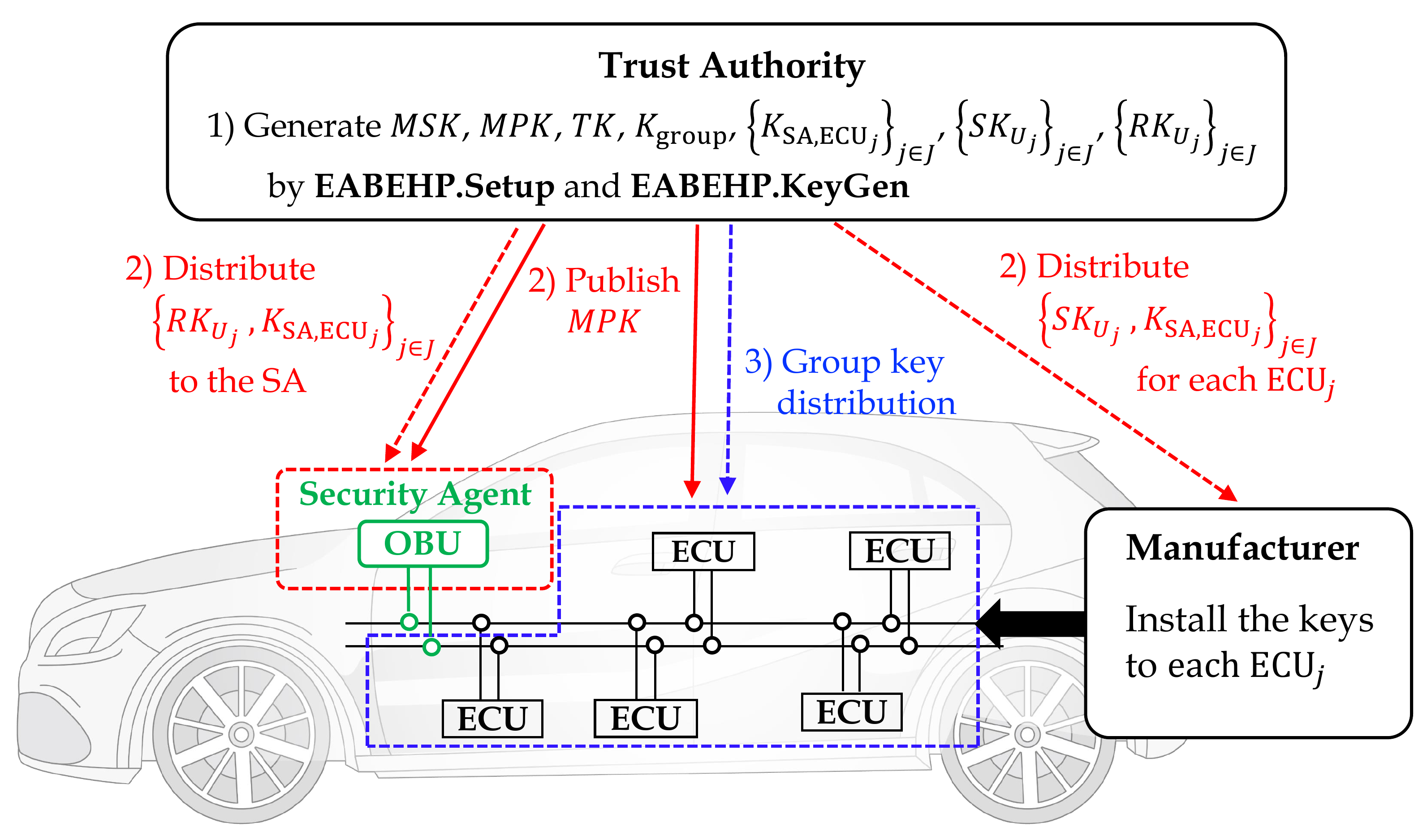}
		}
	\end{center}
\vspace{-3.5mm}
	\caption{
		Long-term secret key management
	}
	\label{fig:Keymanagement}
	\vspace{-3.5mm}
\end{figure}
	
	This section present, the proposed EC-SVC protocol including the key management and the in-vehicle security protocol. In the key management, we present the method to install the symmetric keys as well as the keys in the EABEHP scheme on each device. In addition, we present the in-vehicle security protocol that satisfies the security requirements presented in Sec. II-A. This protocol involves a sender-ECU, receiver-ECUs, and SA, and includes the authentication process between each device, and the EABEHP scheme. We guarantee the security of the protocol by using nonce, signature, symmetric key cryptography, hash function, and EABEHP scheme properly. We first describe the key management and then the in-vehicle security protocol.
	\vspace{-2.5mm}
	\subsection{Key Management}\label{sec:keymanagement}
	
	Figure 2 shows the key management of EC-SVC.
	The trust authority (TA) issues the required cryptographic keys for all entities in the system by the following procedures: (1) TA generates $MSK$, $MPK$, $TK$, $K_\text{group}$, and $\{K_{\text{SA},\text{ECU}_j}\}_{j\in J}$ by \textbf{EABEHP.Setup}. It then generates $SK_{U_j}$ and $RK_{U_j}$ for each $\text{ECU}_j$ by \textbf{EABEHP.KeyGen}. (2) TA publishes $MPK$ and keeps $MSK$ secretly. It then distributes $\{SK_{U_j}, K_{\text{SA},\text{ECU}_j}\}_{j\in J}$ to each user $j$, and $RK_{U_j}$, $TK$ and all $\{K_{\text{SA},\text{ECU}_j}\}_{j\in J}$ to the SA. (3) The TA sends the group key $K_{\text{group}}$ securely to each ECU by executing the group key distribution mechanism [28]. Assume that the size of the security keys is sufficient to provide the security of the system for the life-time of a vehicle.
	\vspace{-2.5mm}
	\subsection{In-vehicle Security Protocol}\label{sec:protocol}

	\begin{figure*}[t!]
		\begin{center}   
			{ 
				\includegraphics[scale=0.4805]{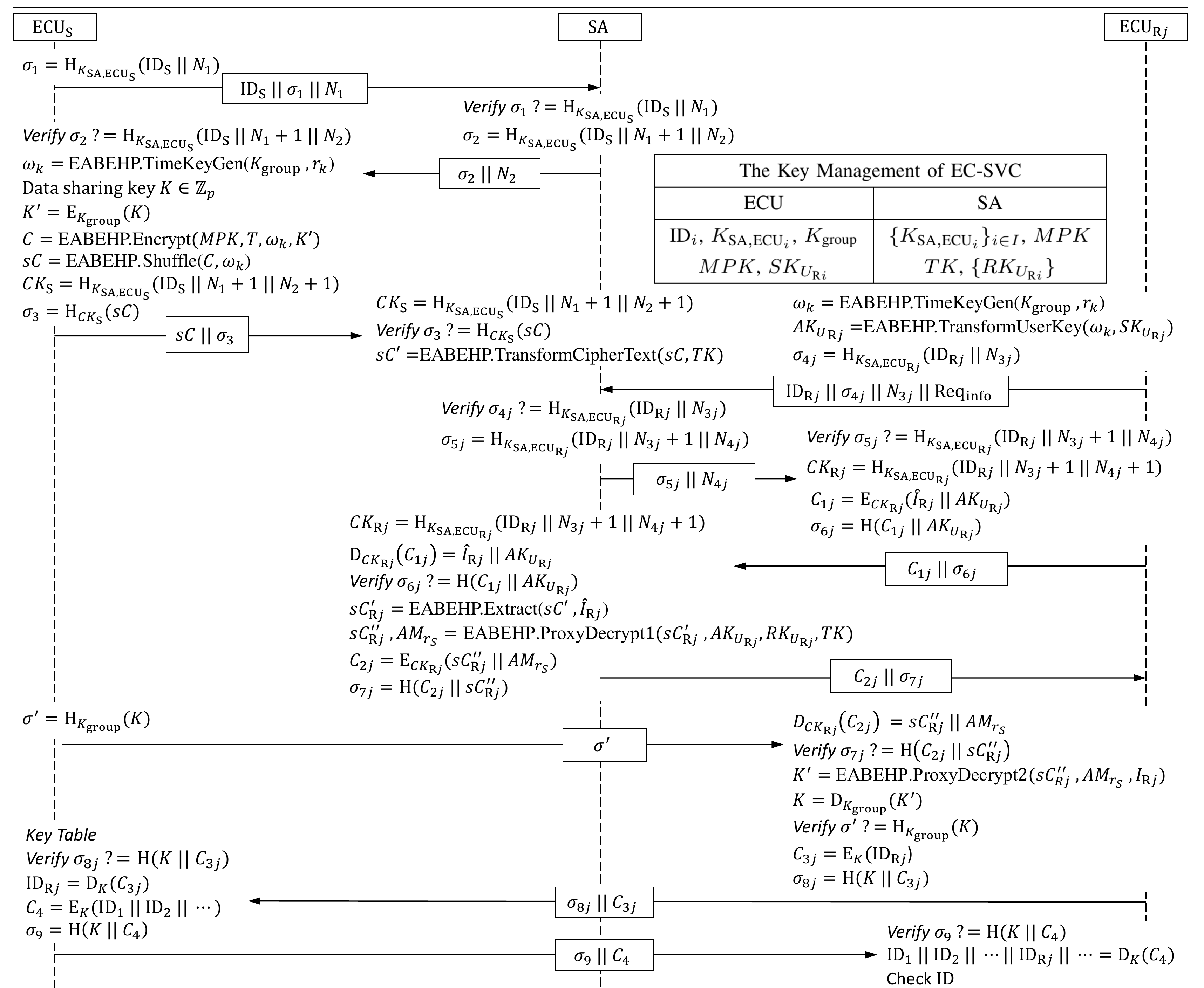}
			}
		\vspace{-2mm}
		\end{center}
		\caption{
			Edge computing-based in-vehicle authenticated key exchange protocol with attribute-based access control
		}
		\label{fig:protocol}
		\vspace{-3.5mm}
	\end{figure*}
	
	This subsection describes the in-vehicle security protocol, which is the edge computing-based in-vehicle authenticated key exchange protocol with attribute-based access control in Fig. 3. 
	The proposed protocol is executed when a vehicle is started and consists of the following twelve steps.
	\begin{enumerate}
	\item The sender-ECU$(\text{ECU}_\text{S})$ generates nonce $N_{1} \in \{0,1\}^{L}$ and ${\sigma}_1=\text{H}_{K_{\text{SA},\text{ECU}_\text{S}}}(\text{ID}_\text{S} || N_1)$. The $\text{ECU}_\text{S}$ then send $\{\text{ID}_\text{S}||\sigma_1||N_1\}$ to the SA.
	
	\item  The SA verifies $\sigma_1$ and generates nonce $N_2 \in \{0,1\}^{L}$. The SA then generates ${\sigma}_2=\text{H}_{K_{\text{SA},\text{ECU}_\text{S}}}(\text{ID}_\text{S}||N_1 +1||N_2)$ and sends $\{\sigma_2||N_2\}$ to the $\text{ECU}_\text{S}$.
	
	\item  The $\text{ECU}_\text{S}$ verifies $\sigma_2$. If passed, the $\text{ECU}_\text{S}$ executes $\omega_k=\text{\textbf{EABEHP.TimeKeyGen}}(K_\text{group}, r_k)$. 
	The $\text{ECU}_\text{S}$ generates data sharing key $K\in {\mathbb{Z}}_q$ and encrypts $K$ as $K'= \text{E}_{K_\text{group}}(K)$. It then computes $C=\text{\textbf{EABEHP.Encrypt}}(MPK, T, \omega_k, K')$. After that, the $\text{ECU}_\text{S}$ computes $sC=\text{\textbf{EABEHP.Shuffle}}(C, \omega_k)$, and generates $CK_\text{S}=\text{H}_{K_{\text{SA},\text{ECU}_\text{S}}}(\text{ID}_\text{S}||N_1 +1||N_2 +1)$ and $\sigma_3=\text{H}_{CK_\text{S}} \allowbreak (sC)$. Subsequently, the $\text{ECU}_\text{S}$ send $\{sC ||\sigma_3\}$ to the SA.
	
	\item  The SA generates $CK_\text{S}=\text{H}_{K_{\text{SA},\text{ECU}_\text{S}}}(\text{ID}_\text{S}||N_1 +1|| N_2 \allowbreak +1)$ and verifies $\sigma_3$. It then computes $sC'=\text{\textbf{EABEHP.TransformCipherText}}(sC, TK)$. After that, the SA waits for the request message from the $\text{ECU}_{\text{R}j}$.
	
	\item  When the SA needs to retrieve the data, sent by $\text{ECU}_{\text{R}j}$, it computes $\omega_k=\text{\textbf{EABEHP.TimeKeyGen}}(K_\text{group}, r_k)$ and $A_{\text{R}j}=\text{\textbf{EABEHP.TransformUserKey}}(\omega_k, SK_{U_{\text{R}j}})$. The $\text{ECU}_{\text{R}j}$ then generates nonce $N_{3j} \in \{0,1\}^{L}$ and $\sigma_{4j}= \text{H}_{K_{\text{SA},\text{ECU}_{\text{R}j}}}(\text{ID}_{\text{R}j} \allowbreak || N_{3j})$. Subsequently, the $\text{ECU}_{\text{R}j}$ sends $\{ID_{\text{R}j} || \sigma_{4j} || N_{3j} || \text{Req}_{\text{info}}\}$ to the SA, where $\text{Req}_\text{info}$ is the message of requesting a process to obtain a data sharing key $K$.
	
	\item  The SA verifies $\sigma_{4j}$ first. If passed, the SA generates nonce $N_{4j} \in \{0,1\}^{L}$ and $\sigma_{5j}= \allowbreak \text{H}_{K_{\text{SA},\text{ECU}_{\text{R}j}}}(\text{ID}_{\text{R}j} || N_{3j}+1 \allowbreak || N_{4j})$. The SA then sends $\{\sigma_{5j} || N_{4j}\}$ to the $\text{ECU}_{\text{R}j}$.
	
	\item  The $\text{ECU}_{\text{R}j}$ verifies $\sigma_{5j}$ and generates $CK_{\text{R}j}= \allowbreak \text{H}_{K_{\text{SA},\text{ECU}_{\text{R}j}}}(\text{ID}_{\text{R}j} || N_{3j}+1 || N_{4j}+1)$.
	The $\text{ECU}_{\text{R}j}$ encrypts $\{\hat{I}_{\text{R}j} || AK_{U_{\text{R}j}}\}$ with the generated $CK_{\text{R}j}$ as $C_{1j}= \text{E}_{CK_{\text{R}j}}(\hat{I}_{\text{R}j} || AK_{U_{\text{R}j}})$ and generates $\sigma_{6j}=\text{H}(C_{1j}||AK_{U_{\text{R}j}})$. Then, the $\text{ECU}_{\text{R}j}$ sends $\{C_{1j}||\sigma_{6j}\}$ to the SA.
	
	\item  After the SA receives $\{C_{1j} || \sigma_{6j}\}$, the SA generates $CK_{\text{R}j}= \allowbreak \text{H}_{K_{\text{SA},\text{ECU}_{\text{R}j}}}(\text{ID}_{\text{R}j} || N_{3j}+1 || N_{4j}+1)$ and then decrypts $C_{1j}$ by $CK_{\text{R}j}$ to obtain the $\{\hat{I}_{\text{R}j} || AK_{U_{\text{R}j}}\}$. The SA verifies $\sigma_{6j}$. The SA then computes $sC'_{\text{R}j}=\text{\textbf{EABEHP.Extract}}(sC', \hat{I}_{\text{R}j})$ and $(sC''_{\text{R}j}, AM_{r_\text{S}})=\text{\textbf{EABEHP.ProxyDecrypt1}}(sC'_{\text{R}j}, AK_{U_{\text{R}j}}, RK_{U_{\text{R}j}}, TK)$, where $r_\text{S}$ is a random number generated by $\text{ECU}_\text{S}$ during the \textbf{EABEHP.Encrypt} process.
	The SA then encrypts $\{sC''_{\text{R}j} || AM_{r_\text{S}}\}$ with $CK_{\text{R}j}$ to generate $C_{2j}$, and generates $\sigma_{7j}=\text{H}(C_{2j}||sC''_{\text{R}j})$. It then sends $\{C_{2j} ||\sigma_{7j}\}$ to the $\text{ECU}_{\text{R}j}$. 
	
	\item  The $\text{ECU}_\text{S}$ sends a $\sigma'=\text{H}_{K_{\text{group}}}(K)$ that allows the $\text{ECU}_{\text{R}}$ to verify that it has received the correct sharing key $K$.
	
	\item The $\text{ECU}_{\text{R}j}$ decrypts $C_{2j}$ to obtain $\{sC''_{\text{R}j} || AM_{r_\text{S}}\}$ and verifies $\sigma_{7j}=\text{H}(C_{2j}||sC''_{\text{R}j})$. The $\text{ECU}_{\text{R}j}$ then computes $K'=\text{\textbf{EABEHP.ProxyDecrypt2}}(sC''_{\text{R}j} , AM_{r_\text{S}}, I_{\text{R}j})$, and decrypts it with $K_{\text{group}}$ to obtain $K$. Afterwards, the $\text{ECU}_{\text{R}j}$ verifies signature $\sigma'$. Finally, the $\text{ECU}_{\text{R}j}$ generates $C_{3j}=\text{E}_K(\text{ID}_{\text{R}j})$ and $\sigma_{8j}=\text{H}(K||C_{3j})$ and sends $\{\sigma_{8j}||C_{3j}\}$ to the $\text{ECU}_\text{S}$.
	
	\item  The $\text{ECU}_{\text{S}}$ stores the ID of the $\text{ECU}_{\text{R}j}$ that exchanged the data sharing key in the key table. The $\text{ECU}_{\text{S}}$ verifies $\sigma_{8j}$ and decrypts $C_{3j}$ by $K$ to obtain $\text{ID}_{\text{R}j}$. The $\text{ECU}_{\text{S}}$ then generates $C_4=\text{E}_K(\text{ID}_1||\text{ID}_2||\cdots)$ for the encryption of all the $\text{ECU}_{\text{R}j}$'s identities and $\sigma_9=\text{H}(K||C_4)$. Subsequently, the $\text{ECU}_\text{S}$ sends $\{\sigma_9||C_4\}$ to the $\text{ECU}_{\text{R}j}$.
	
	\item  The $\text{ECU}_{\text{R}j}$ verifies signature $\sigma_9$ and decrypts $C_4$. If the $\text{ECU}_{\text{R}j}$ can find its own identity contained, it can authenticate the $\text{ECU}_{\text{S}}$ successfully, which verified the mutual authentication. 
	\\
	\end{enumerate}
\vspace{-4.5mm}
	\section{Security Analysis}\label{sec:securityanalysis}
	
	This section proves the security of the cryptographic scheme proposed by Sec. III-B, and proves the security of the proposed protocol based on the security definition and model of Sec. II-C.
	\vspace{-3.5mm}
	\subsection{Security Analysis of Enhanced Attribute-based Encryption with Hidden Policy and Credential}\label{sec:analysisEABEHP}
	
	\begin{theorem}[Confidentiality of EABEHP]
		The proposed EABEHP is with ciphertext indistinguishability against chosen plaintext attack and restricted user coalition attack~(C-IND-CPA-RUCA) if the decisional Diffie-Hellman~(DDH) assumption holds.
	\end{theorem}
		\noindent {\bf Proof Sketch.} 
		The proof of C-IND-CPA-RUCA for the proposed EABEHP consists of two parts. One is the confidentiality of the produced ciphertext, i.e., $C$, in EABEHP. Moreover, $sC'$ and $sC'_{r}$ are also considered as the ciphertext of EABEHP since they are transformed from $C$ and the only difference is that the positions of tuples of ciphertext are shuffled. Thus, the C-IND-CPA-RUCA security can be proven according to the same security proof for the proposed PEAPOD in [22] since the structure of the ciphertext in EABEHP is the same as that in PEAPOD. The other is the confidentiality of $C$, $sC'$, $sC'_{r}$, and $sC''_{r}$ against the proxy for \textbf{ProxyDecrypt1}. The second part of the proof of C-IND-CPA-RUCA considers that the confidentiality is guaranteed against the proxy even though $AK_{U}$ and $RK_{U}$ are known to the proxy. $RK_{U}$ is the re-encryption key and it preserves the same structure of that in the PEAPOD scheme. Thus, the exposure of $RK_{U}$ will not affect the confidentiality of the ciphertext in EABEHP. In addition, the exposure of $AK_{U}$ will not affect the confidentiality as well since additional secret key $\omega_{k}$, unknown to the proxy, is introduced to protect the secrets, $a_{j,i}$, contained in $AK_{U}$. Thus, without known $\omega_k$, the proxy cannot eliminate the factor, $g^{\omega_k}$, of blinding the message in the ciphertext to break the confidentiality of EABEHP.
	
	\begin{theorem}[Policy Privacy of EABEHP]
		Policy privacy holds if no one, including the security agent can learn any knowledge of the given policy $T$ for encryption in the proposed EABEHP scheme.
	\end{theorem}
	\noindent {\bf Proof Sketch.} In EABEHP, a message $M$ to be encrypted will first be encoded as $p_i$ for $i\in{}I$ such that $\Pi_{t_i=1 \wedge i\in{}I}p_i=M$. Here, $T=\{t_i\}_{i\in I}$ is a policy set, where $t_i = 1, 0, -1$ if the attribute $i$ is required, irrelevant, and unrequired respectively. Here, $p_i = 1$ when $t_i = 0$ and $p_i=\alpha_i$ when $t_i = -1$, for randomly selected $\alpha_i \in {\mathbb{Z}}_q$. Afterwards, one can then encrypt each $p_i$ as $(A,B_i,C)$ with the public key of its corresponding attribute $i$. Thus, the only way to learn the given policy $I$ depends on the generated $p_i$. However, each $p_i$ is encrypted using ElGamal encryption which is the primitive of EABEHP with indistinguishability under chosen plaintext attack (IND-CPA) security. Thus, no one can learn any knowledge from $p_i$ by the ciphertext $(A,B_i,C)$. From the above, no one, including the security agent, can learn the knowledge of policy. Consequently, EABEHP is with policy privacy.
	
	\begin{theorem}[Credential Privacy of EABEHP]
		The proposed EABEHP is with hidden credentials against outsider and decryption proxy if the underlying hash function is a pseudorandom function, and the shuffle function is a pseudorandom permutation.
	\end{theorem}
	\noindent {\bf Proof Sketch.} Since the attribute information, i.e., $AK_{U_r}$ or $\hat{I}_r$, of each user will be exposed when during the execution of \textbf{ProxyDecrypt1} or \textbf{Extract} function, the privacy of user credential is guaranteed if the original attribute information cannot be disclosed. (1) $AK_{U_r}$ is a combination of $\omega_k$ and user secret keys for each attribute generated by the transformuserkey algorithm, where $\omega_k=\text{H}(r_k||K_\text{group})$, and $K_\text{group}$ is a pre-distributed key to legitimate users in the system. Therefore, no one has non-negligible probability to distinguish $AK_{U_r}$ by distinguishing a $\omega_k$ from random string based on the security of pseudorandom function. (2) $\hat{I}_r$ is the set of inverse permuted attribute indices from the set of original attribute indices by a pseudorandom permutation, shuffle function $SH$ with a given $\omega_k$, which is only known between users. Thus, no one has non-negligible probability to distinguish a permuted index from an original index based on the security of pseudorandom permutation. Thus, the proposed EABEHP is with hidden credentials based on the security of the pseudorandom function and permutations.
\vspace{-2.5mm}
	\subsection{Security Analysis of EC-SVC Protocol}\label{sec:analysisprotocol}
	In this subsection, we present to the security analysis of the protocol proposed in Sec. IV-B. 	
	The proposed protocol proves that mutual authentication, attribute-based key exchange, policy privacy, and credential privacy have been achieved as follows. 
	
	\begin{theorem}[EC-SVC Security] 
		The proposed EC-SVC protocol is said to be the attribute-based authenticated key exchange protocol with hidden policy and credential if H is a pseudorandom function, $\text{E}_\text{S}$ is a pseudorandom permutation, and EABEHP is a C-IND-CPA-RUCA-secure and P-IND-CPA-UCA-secure attribute-based encryption scheme.
	\end{theorem}
	The advantage $\text{Adv}_{\mathcal{A}}^{\text{EC-SVC}}$ that an attacker $\mathcal{A}$ break the security of EC-SVC protocol are given by
	\begin{align} \notag
	\text{Adv}_{\mathcal{A}}^{\text{EC-SVC}} \leq &11\text{Adv}_{\text{H}}+5\text{Adv}_{\text{E}_\text{S}}+2{\text{Adv}_{\text{C-IND-CPA}}}\\
	&+2{\text{Adv}_{\text{P-IND-CPA}}}.
	\end{align}
	where $\text{Adv}_{\text{H}}$ is an advantage that breaks the security of the pseudorandom function, $\text{Adv}_{\text{E}_\text{S}}$ is an advantage that breaks the security of the pseudorandom permutation, $\text{Adv}_{\text{C-IND-CPA}}$ is an advantage that breaks the security of the C-IND-CPA-RUCA security of the EABEHP, and $\text{Adv}_{\text{P-IND-CPA}}$ is an advantage that breaks the security of the P-IND-CPA-UCA security of the EABEHP.
	
	We proceed with the security game to prove the security of the proposed protocol. The security game proceeds each four requirements mentioned above and claims that the advantages of $\mathcal{A}$ for the proposed protocol can be negligible, depending on the advantages of $\mathcal{A}$ in each game, where $\mathcal{A}$ is an attacker that breaks the security of mutual authentication, attribute-based key exchange, policy privacy, and credential privacy. 
	We denote $\text{Adv}_{\mathcal{A},i}^{\text{EC-SVC}}$ as the advantage of $\mathcal{A}$ in game $G_i$.
	
	\textit{Game $G_0$} : This is a real game, $\mathcal{A}$ has access to EABEHP's master public key $MPK$, all ECU's identity (ID) $\{ID_i\}_{i=0,1,\ldots}$. In addition, $\mathcal{A}$ has the ability to query all oracles specified in Sec. II-C and knows all the structure of the protocol. Since this paper has shown that EABEHP can be proven IND-CPA secure by simulating EABEHP in Sec. V-A, all the parameters related to EABEHP can be successfully simulated. Therefore, we have
	\begin{align}
	\text{Adv}_{\mathcal{A}}^{\text{EC-SVC}}=\text{Adv}_{\mathcal{A},0}^{\text{EC-SVC}}.
	\end{align}
	\textit{Game $G_1$} (Mutual Authentication).
	In the game $G_1$, We describe the events of the game as follows. $\text{E}_1$ is an event in which $\mathcal{A}$ impersonates $\text{ECU}_\text{S}$ by sending the correct $\sigma_1$ to the SA. 
	The $\text{E}_2$ is an event in which $\mathcal{A}$ impersonates the SA by sending the correct $\sigma_2$ to the $\text{ECU}_\text{S}$.
	The $\text{E}_3$ is an event in which $\mathcal{A}$ impersonates the $\text{ECU}_{\text{R}j}$ by sending the correct $\sigma_{4j}$ to the SA. 
	The $\text{E}_4$ is an event in which $\mathcal{A}$ impersonates the SA by sending the correct $\sigma_{5j}$ to $\text{ECU}_{\text{R}j}$. 
	The $\text{E}_5$ is an event in which $\mathcal{A}$ impersonates the $\text{ECU}_{\text{R}j}$ by sending the correct $\sigma_{8j} || C_{3j}$ to the $\text{ECU}_\text{S}$.
	The $\text{E}_6$ is an event in which $\mathcal{A}$ impersonates the $\text{ECU}_\text{S}$ by sending the correct $\sigma_9 || C_4$ to the $\text{ECU}_{\text{R}j}$.
	We construct a simulator $S_1$ of the EC-SVC that interacts with $\mathcal{A}$ as the security game defined in Definition 4. In addition, $S_1$ is provided with the master public key of EABEHP to successfully simulate EC-SVC. If the $\text{E}_1$ happens, $S_1$ can exploit the ability of $\mathcal{A}$ to break the underlying pseudorandom function security. Hence, we have
	\begin{align} \notag
	&\text{Adv}_{\text{H}} \geq \{\text{Pr}[\text{S}_{\text{H}},\text{E}_1]+\text{Pr}[\text{S}_{\text{H}},\neg\text{E}_1]\}-\frac{1}{2}\\ \notag
	&= \{\text{Pr}[\text{S}_{\text{H}}|\text{E}_1]\times\text{Pr}[\text{E}_1]+\text{Pr}[\text{S}_{\text{H}}|\neg\text{E}_1]\times(1-\text{Pr}[\text{E}_1]\}-\frac{1}{2}\\ 
	&=
	\{1\times\text{Adv}_{\text{E}_1}+\frac{1}{2}\times(1-\text{Adv}_{\text{E}_1})\}-\frac{1}{2} = \frac{\text{Adv}_{\text{E}_1}}{2},
	\end{align}
	where $\text{S}_{\text{H}}$ is the event of distinguishing a pseudorandom function from a truly random function successfully, the $\neg\text{E}_1$ is the complementary event of the $\text{E}_1$, $\text{Adv}_{\text{E}_1}$ is the advantage of the $\text{E}_1,$ which is the probability that an attacker sends a valid $\sigma_1$ to impersonate an $\text{ECU}_\text{S}$. Therefore, we have $\text{Adv}_{\text{E}_1}\leq 2\text{Adv}_{\text{H}}$. For the probabilities of events $\text{E}_2$, $\text{E}_3$, and $\text{E}_4$, we have $\text{Adv}_{\text{E}_2}\leq 2\text{Adv}_{\text{H}}$, $\text{Adv}_{\text{E}_3}\leq 2\text{Adv}_{\text{H}}$, and $\text{Adv}_{\text{E}_4}\leq 2\text{Adv}_{\text{H}}$. The security analysis regarding $\text{E}_5$ can be divided into two cases: (1) When $\text{E}_5$ happened, $S_1$ can also break the security of underlying pseudorandom function or pseudorandom permutation by exploiting the ability of $\mathcal{A}$. Thus, we have $\text{Adv}_{\text{E}_{5}} \leq{} 2\text{Adv}_{\text{E}_\text{S}}$, when $\text{S}_1$ simulates the protocol based on the function, which is either a pseudorandom permutation or a random permutation. In addition, when $\text{S}_1$ simulates the protocol based on the function, which is either a pseudorandom function or a random function, we have $\text{Adv}_{\text{E}_{5}} \leq{} 2\text{Adv}_{\text{H}}$. From the above, we have
	\begin{align}
	\text{Adv}_{\text{E}_{5}} \leq{} \text{Adv}_{\text{E}_{\text{S}}} + \text{Adv}_{\text{H}}.
	\end{align}
	(2) When $\text{E}_5$ happened, $S_1$ can also break the security of underlying pseudorandom permutation or C-IND-CPA-RUCA by exploiting the ability of $\mathcal{A}$. Thus, we have
	\begin{align}
	\text{Adv}_{\text{E}_{5}} \leq{} \text{Adv}_{\text{E}_{\text{S}}} + \text{Adv}_{\text{C-IND-CPA}}.
	\end{align}
	Through the results of both cases, we have
	\begin{align}
	\text{Adv}_{\text{E}_{5}} \leq{} \text{Adv}_{\text{E}_{\text{S}}} + \frac{1}{2}(\text{Adv}_{\text{H}}+\text{Adv}_{\text{C-IND-CPA}}).
	\end{align}
	In the same way, we have $\text{Adv}_{\text{E}_6} \leq{} \text{Adv}_{\text{E}_{\text{S}}} + \frac{1}{2}(\text{Adv}_{\text{H}}+\text{Adv}_{\text{C-IND-CPA}})$. Finally, we have
	\begin{align}\label{eq:gameg1}
	\text{Adv}_{\mathcal{A},0}^{\text{EC-SVC}}\hspace{-1mm} \leq &\text{Adv}_{\mathcal{A},1}^{\text{EC-SVC}}\hspace{-1mm}+\hspace{-0.8mm}9\text{Adv}_{\text{H}}\hspace{-0.8mm}+\hspace{-0.8mm}2\text{Adv}_{\text{E}_\text{S}}\hspace{-1mm}+\hspace{-0.8mm}\text{Adv}_{\text{C-IND-CPA}}
	\end{align}
	
	\textit{Game $G_2$} (Attribute-based key exchange).
	The proposed protocol achieves attribute-based key exchange through the \textit{enhanced attribute-based encryption with hidden policy and credential} newly proposed in Sec. III-B. In game $G_2$, we construct a simulator $S_2$ that interacts with $\mathcal{A}$ in the security games defined in Definition 5. $S_2$ is provided with the master public key of EABEHP to successfully simulate EC-SVC. $\mathcal{A}$ queries the \textbf{Test} after interacting with the security game with $S_2$. $S_2$ responds to the $\mathcal{A}$ with an attribute-based key $K$ or a random string according to a random bit. If $\mathcal{A}$ can successfully guess the attribute-based key $K$, $S_2$ can also break the security of underlying pseudorandom permutation or C-IND-CPA-RUCA by exploiting the ability of $\mathcal{A}$. Therefore, we have
	\begin{align} \notag
	&\text{Adv}_{\text{E}_\text{S}} \geq \{\text{Pr}[\text{S}_{\text{E}_\text{S}},\text{E}_\text{AKE}]+\text{Pr}[\text{S}_{\text{E}_\text{S}},\neg\text{E}_\text{AKE}]\}-\frac{1}{2}\\ \notag
	&= \{\text{Pr}[\text{S}_{\text{E}_\text{S}}|\text{E}_\text{AKE}]\times\text{Pr}[\text{E}_\text{AKE}] \\
	&\quad
	+\text{Pr}[\text{S}_{\text{E}_\text{S}}|\neg\text{E}_\text{AKE}] \times(1-\text{Pr}[\text{E}_\text{AKE}]\}-\frac{1}{2}\\ \notag
	&=
	\{1\times\text{Adv}_{\text{E}_\text{AKE}}+\frac{1}{2}\times(1-\text{Adv}_{\text{E}_\text{AKE}})\}-\frac{1}{2}=\frac{\text{Adv}_{\text{E}_\text{AKE}}}{2},
	\end{align}
	where $\text{Adv}_{\text{E}_\text{AKE}}$, which is the advantage of the $\text{E}_\text{AKE}$, is probability that an attacker distinguishes the attribute based key $K$ from a random string. Therefore, we have $\text{Adv}_{\text{E}_\text{AKE}} \leq 2\text{Adv}_{\text{E}_\text{S}}$. Similar to the game $G_1$, we have $\text{Adv}_{\text{E}_\text{AKE}} \leq 2\text{Adv}_{\text{C-IND-CPA}}$. From the above, we have
	\begin{align}
	\text{Adv}_{\text{E}_\text{AKE}} \leq \text{Adv}_{\text{E}_{\text{S}}}+\text{Adv}_{\text{C-IND-CPA}}.
	\end{align}
	Therefore, we have
	\begin{align}\label{eq:gameg2}
	\text{Adv}_{\mathcal{A},1}^{\text{EC-SVC}} &\leq \text{Adv}_{\mathcal{A},2}^{\text{EC-SVC}}+\text{Adv}_{\text{E}_{\text{S}}}+\text{Adv}_{\text{C-IND-CPA}}.
	\end{align}
	
	\textit{Game $G_3$} (Policy Privacy).
	The policy privacy of the proposed protocol can be analyzed in a similar way to the game $G_2$. In game $G_3$, we construct a simulator $S_3$ that interacts with $\mathcal{A}$ in the security games defined in Definition 6. $S_3$ is provided with the master public key of EABEHP to successfully simulate EC-SVC. $\mathcal{A}$ queries the \textbf{TestPolicy} after interacting with the security game with $S_3$. $S_3$ responds to the $\mathcal{A}$ with an $P_0$ or $P_1$ according to a random bit. If $\mathcal{A}$ can successfully guess the correct policy, then $\mathcal{A}$ has the advantage of breaking P-IND-CPA-UCA security of EABEHP. Therefore, we have
	\begin{align} \notag
	&\text{Adv}_{\text{P-IND-CPA}} \\ \notag
	&\geq  \{\text{Pr}[\text{S}_{\text{P-IND-CPA}},\text{E}_\text{PP}]+\text{Pr}[\text{S}_{\text{P-IND-CPA}},\neg\text{E}_\text{PP}]\}-\frac{1}{2}\\ \notag
	&= \{\text{Pr}[\text{S}_{\text{P-IND-CPA}}|\text{E}_\text{PP}]\times\text{Pr}[\text{E}_\text{PP}] \\ \notag 
	&\quad
	+\text{Pr}[\text{S}_{\text{P-IND-CPA}}|\neg\text{E}_\text{PP}] \times(1-\text{Pr}[\text{E}_\text{PP}]\}-\frac{1}{2}\\ 
	&=
	\{1\times\text{Adv}_{\text{PP}}+\frac{1}{2}\times(1-\text{Adv}_{\text{PP}})\}-\frac{1}{2} = \frac{\text{Adv}_{\text{PP}}}{2},
	\end{align}
	where $\text{E}_\text{PP}$ is the event that distinguishing the correct policy from random string with additional advantage, and $\text{Adv}_{\text{PP}}$ is the advantage of breaking policy privacy. Thus, we have
	\begin{align}\label{eq:gameg3}
	\text{Adv}_{\mathcal{A},2}^{\text{EC-SVC}} &\leq \text{Adv}_{\mathcal{A},3}^{\text{EC-SVC}}+2{\text{Adv}_{\text{P-IND-CPA}}}.
	\end{align}

	\begin{table}[] \scriptsize
		\caption{Comparison on Security with Related works}
		\begin{center}
		\renewcommand{\arraystretch}{1.4}
		\begin{tabular}{c|cccc|c}
			\multirow{2}{*}{} & [9] & [11] & [12] & [13] & \multirow{2}{*}{\begin{tabular}[c]{@{}c@{}}Our work \\ (EC-SVC)\end{tabular}} \\
			& [14] & [16] & [17] & [18]  &                  \\ \cline{1-1} \cdashline{2-6}
			\multirow{2}{*}{Mounted Additional Device} & $\surd$ & $\surd$ & $\surd$ & $\times$ & \multirow{2}{*}{$\times$} \\
			& $\times$ & $\times$ & $\times$ & $\times$ &                   \\ \cline{1-1} \cdashline{2-6}
			\multirow{2}{*}{\begin{tabular}[c]{@{}c@{}}Message \\ Authentication and Integrity\end{tabular}} & $\surd$ & $\surd$ & $\surd$ & $\surd$ & \multirow{2}{*}{$\surd$} \\
			& $\surd$ & $\surd$ & $\surd$ & $\surd$ &                   \\ \cline{1-1} \cdashline{2-6}
			\multirow{2}{*}{Data Confidentiality} & $\surd$ & $\times$ & $\times$ & $\times$ & \multirow{2}{*}{$\surd$} \\
			& $\times$ & $\surd$ & $\surd$ & $\surd$ &                   \\ \cline{1-1} \cdashline{2-6}
			\multirow{2}{*}{\begin{tabular}[c]{@{}c@{}}Resistance \\ to Replay Attacks\end{tabular}} & $\surd$ & $\times$ & $\times$ & $\surd$ & \multirow{2}{*}{$\surd$} \\
			& $\surd$ & $\surd$ & $\surd$ & $\surd$ &                   \\ \cline{1-1} \cdashline{2-6}
			\multirow{2}{*}{\begin{tabular}[c]{@{}c@{}}Attribute-based\\ Access Control\end{tabular}} & $\times$ & $\times$ & $\times$ & $\times$ & \multirow{2}{*}{$\surd$} \\
			& $\times$ & $\times$ & $\times$ & $\times$ &                   \\ \cline{1-1} \cdashline{2-6}
			\multirow{2}{*}{\begin{tabular}[c]{@{}c@{}}Privacy Preserving \\ for Corrupted Devices\end{tabular}} & - & - & - & - & \multirow{2}{*}{$\surd$} \\
			& - & - & - & - & \\ \hline
		\end{tabular}
	\end{center}
	\label{table:comparison}
	\vspace{-6mm}
	\end{table}

	\textit{Game $G_4$} (Credential Privacy).
	In game $G_4$, we construct a simulator $S_4$ that interacts with $\mathcal{A}$ in the security games defined in Definition 7. $S_4$ successfully simulates EC-SVC with the supplied EABEHP's master public key. After interacting with the security game with $S_4$, $\mathcal{A}$ queries the \textbf{TestCert}. $S_4$ responds to $\mathcal{A}$ with a target credential or randomly selected credential according to a random bit. $\text{E}_9$ is an event in which credentials are successfully guessed in the \textbf{ProxyDecrypt1} algorithm by adversary $\mathcal{A}$. $\text{E}_{10}$ is a case in which the credentials are successfully guessed in the \textbf{Extract} algorithm by $\mathcal{A}$.
	If the $\text{E}_9$ happens, $\mathcal{A}$ has the advantage of breaking pseudorandom function security. Therefore, we have
	\begin{align} \notag
	&\text{Adv}_{\text{H}} \geq  \{\text{Pr}[\text{S}_{\text{H}},\text{E}_9]+\text{Pr}[\text{S}_{\text{H}},\neg\text{E}_9]\}-\frac{1}{2}\\ \notag
	&= \{\text{Pr}[\text{S}_{\text{H}}|\text{E}_9]\times\text{Pr}\text{E}_9]+\text{Pr}[\text{S}_{\text{H}}|\neg\text{E}_9]\times(1-\text{Pr}[\text{E}_9]\}-\frac{1}{2}\\
	&=
	\{1\times\text{Adv}_{\text{E}_9}+\frac{1}{2}\times(1-\text{Adv}_{\text{E}_9})\}-\frac{1}{2} = \frac{\text{Adv}_{\text{E}_9}}{2},
	\end{align}
	where $\text{Adv}_{\text{E}_9}$, which is the advantage of the $\text{E}_9$, is the probability that an attacker distinguishes the real user private key from a random string. Therefore, we have $\text{Adv}_{\text{E}_9}\leq 2\text{Adv}_{\text{H}}$. In the same way, for $\text{E}_{10}$ we have $\text{Adv}_{\text{E}_{10}}\leq 2\text{Adv}_{\text{E}_\text{S}}$. Therefore, we have $\text{Adv}_{\text{CP}} \leq 2\text{Adv}_{\text{H}}+2\text{Adv}_{\text{E}_\text{S}}$, where $\text{E}_\text{CP}$ is the event that distinguishing the target credential from randomly selected credential, and $\text{Adv}_{\text{CP}}$ is the advantage of breaking credential privacy.
	\begin{align}\label{eq:gameg4}
	\text{Adv}_{\mathcal{A},3}^{\text{EC-SVC}} &\leq \text{Adv}_{\mathcal{A},4}^{\text{EC-SVC}}+2\text{Adv}_{\text{H}}+2{\text{Adv}_{\text{E}_\text{S}}}.
	\end{align}

	There are no additional advantages beyond those analyzed in the game above. Thus, by equations (14), (17), (19), and (21) we can claim that the advantages of $\mathcal{A}$ to the proposed EC-SVC are as given by
	\begin{align} \notag
	\text{Adv}_{\mathcal{A}}^{\text{EC-SVC}} \leq &11\text{Adv}_{\text{H}}+5\text{Adv}_{\text{E}_\text{S}}+2{\text{Adv}_{\text{C-IND-CPA}}}\\
	&+2{\text{Adv}_{\text{P-IND-CPA}}}.
	\end{align}
	Finally, the overall security comparison between security protocols and related works is shown in Table I. The work satisfies all the security requirements without mounting additional components on ECUs.
	\vspace{-2.5mm}

	\section{Performance Analysis}\label{sec:performanceanalysis}
	In this section, we evaluate the performance in various aspects for demonstrating that the proposed security protocol is practical in in-vehicle scenarios. This work builds up the testbed based on the hardware and software which are the Raspberry Pi, TMS320C28346, and CANoe by Vector Co [29]. 
	Unless otherwise specified, the simulation environment in Fig. 4 and the specifications of the equipment in Table II are used. The testbed adopts CANoe to implement in-vehicle network based on the flexible data rate (CAN-FD) standard [30]. 
	\begin{figure}[t!]
		\begin{center}   
			{ 
				\includegraphics[scale=0.31]{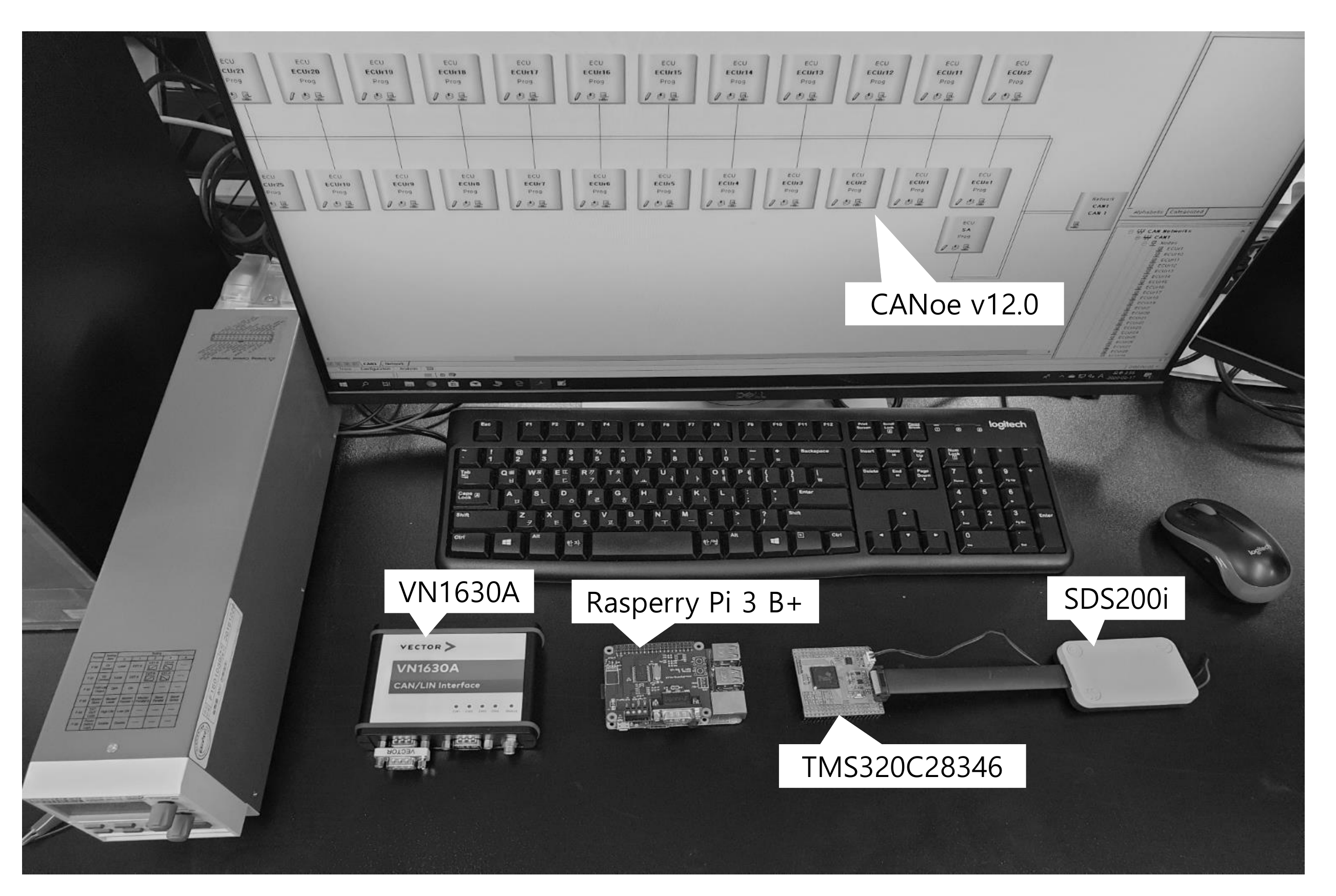}
			}
		\end{center}
		\vspace{-5mm}
		\caption{
			Performance evaluation environment.
		}
		\label{fig:environment}
		\vspace{-4mm}
	\end{figure}
	\begin{table} \scriptsize
		\caption{Hardware and Software for Performance Evaluation}
		\begin{center}
			\renewcommand{\arraystretch}{1.4}
			\begin{tabular} {c|c}
				\hline 
				{\bf Model} & {\bf Note}
				\\
				\hline
				\begin{tabular}[c]{@{}l@{}}Raspberry Pi 3 B+ \\ (Single-board Computer) \end{tabular} & \begin{tabular}[c]{@{}l@{}}Clock speed \\ : 1.4GHz or 600MHz \end{tabular} \\ \hline
				\begin{tabular}[c]{@{}l@{}}TI TMS320C28346 \\ (Micro controller unit (MCU)) \end{tabular} & Clock speed : 300MHz \\ \hline
				SDS200i & JTAG Emulator\\ \hline
				VN1630A & CAN-FD Network Interface\\ \hline
				Java JCA/JCE & Java Cryptography Package\\ \hline
				Code Composer Studio V9.3 & MCU Compiler\\ \hline
				CANoe V12.0 & In-vehicle Network simulator\\ \hline
			\end{tabular}
		\end{center}
		\label{table:specification}
		\vspace{-5mm}
	\end{table}
	This section first analyzes the execution time of each cryptographic algorithm for each device. We then evaluate the performance of the proposed security protocols in the simulation environment implemented.
	
\vspace{-2.5mm}
	\subsection{Cryptographic Algorithm Evaluation}\label{sec:algorithmevaluation}
	\vspace{-0.5mm}
	
		\begin{table} \scriptsize
		\caption{Execution time of Cryptographic algorithm}
		\begin{center}
			\renewcommand{\arraystretch}{1.4}
			\begin{tabular} {c|c|c|c}
				\hline
				& \multicolumn{3}{c}{{\bf Algorithm execution time $(\mu s)$}} \\
				\hline 
				{\bf Algorithm} & {\bf SHA-256} & {\bf AES128(Enc)} & {\bf AES128(Dec)}
				\\ \hline
				ECU & 130.8 & 149.5 & 198.9\\ \hline
				SA (600MHz) & 8.4 & 5.4 & 6.7\\ \hline
				SA (1.4GHz) & 3.6 & 12.7 & 13.8\\ \hline
			\end{tabular}
		\end{center}
		\label{table:basicalgorithmtime}
		\vspace{-5mm}
	\end{table}

	\begin{table}[t!] \scriptsize
		\caption{Execution time of EABEHP Algorithm}
		\begin{center}
			\renewcommand{\arraystretch}{1.4}
			\begin{tabular}{c|c|c|c|c|c}
				\hline
				\multicolumn{2}{c|}{} & \multicolumn{4}{c}{\textbf{Algorithm execution time (ms)}} \\  \hline
				\multicolumn{2}{c|}{\multirow{2}{*}{\begin{tabular}[c]{@{}c@{}}\textbf{Number of}\\ \textbf{system attributes}\end{tabular}}} & 4 & 8 & 12 & 16 \\ \cline{3-6}
				\multicolumn{2}{c|}{} & 20 & 24 & 28 & 32 \\  \hline
				\multirow{2}{*}{\begin{tabular}[c]{@{}c@{}}EABEHP\\ Encrypt+Shuffle\end{tabular}} & \multirow{2}{*}{ECU} & 144.7 & 241.1 & 338.8 & 436.9 \\  \cline{3-6}
				&                   & 529.5 & 635.5 & 714.8 & 817.9 \\  \hline
				\multirow{4}{*}{\begin{tabular}[c]{@{}c@{}}EABEHP\\ TransformCipherText\end{tabular}} & \multirow{2}{*}{\begin{tabular}[c]{@{}c@{}}SA\\ (600MHz)\end{tabular}} & 7 & 13 & 20.9 & 27.8 \\  \cline{3-6}
				&                  & 34.4 & 41.8 & 47.6 & 54.8 \\  \cline{2-6}
				& \multirow{2}{*}{\begin{tabular}[c]{@{}c@{}}SA\\ (1.4GHz)\end{tabular}} & 3 & 6 & 9 & 12 \\  \cline{3-6}
				&                   & 14.5 & 17.5 & 21.2 & 23.6 \\  \hline \hline
				\multicolumn{2}{c|}{\multirow{2}{*}{\begin{tabular}[c]{@{}c@{}}\textbf{Number of}\\ \textbf{receiver attributes}\end{tabular}}} & 4 & 8 & 12 & 16 \\  \cline{3-6}
				\multicolumn{2}{c|}{} & 20 & 24 & 28 & 32 \\  \hline
				\multirow{4}{*}{\begin{tabular}[c]{@{}c@{}}EABEHP\\ Extract+ProxyDecrypt1\end{tabular}} & \multirow{2}{*}{\begin{tabular}[c]{@{}c@{}}SA\\ (600MHz)\end{tabular}} & 1.92 & 2.05 & 2.25 & 2.46 \\  \cline{3-6}
				&  & 2.65 & 3 & 3.24 & 3.64 \\  \cline{2-6}
				& \multirow{2}{*}{\begin{tabular}[c]{@{}c@{}}SA\\ (1.4GHz)\end{tabular}} & 0.82 & 0.89 & 0.96 & 1.08 \\  \cline{3-6}
				&                   & 1.12 & 1.25 & 1.44 & 1.56 \\ \hline
			\end{tabular}
		\end{center}
		\label{table:EABEHPalgorithmtime}
		\vspace{-5.5mm}
	\end{table}
	This subsection evaluates the execution time of the different cryptographic algorithms (i.e., SHA-256, AES-128, EABEHP). The cryptographic algorithm is implemented and measured in Java Cryptography Architecture (JCA) / Java Cryptography Extension (JCE) and Code Composer Studio. We construct the EABEHP algorithm based on ElGamal encryption. In addition, for better accuracy, we measure the execution time for 10,000 times repetitively and obtain the average execution time. We use the TMS320C28346 as the ECU and the Raspberry Pi as the SA.
	
	In the proposed security protocol, 48byte input is used to the SHA256 algorithm, and 16byte output is obtained by truncated MAC [16], [17], [31]. In addition, 48bytes out of 64bytes in the data payload is used as input data to AES128 algorithm. The length of all messages sent by the ECU and the SA is 64bytes, and the remnant is assumed to be padded.
	Under this description, the measured execution time of various cryptographic algorithm is shown in Table III. 
	Table III show the cryptographic algorithm execution time of SHA-256, AES128 Encryption, and AES128 Decryption at the ECU and the SA.
	
	Table IV shows the EABEHP Encrypt and Shuffle algorithm execution time at the ECU\footnote{Note that, it was not possible to perform EABEHP Encrypt and Shuffle at the ECU due to the hardware limitation. Hence, we obtain the execution time of them by measuring of Raspberry Pi and then scaling the time by the execution time ratio of other operations (e.g., SHA-256 and AES128 in Table IV). We expect that the ECU, developed in the near future, will be able to support the advanced cryptographic operations.}, and the EABEHP TransformCipherText execution time and EABEHP Extract and ProxyDecrypt1 execution time at the SA.
	The EABEHP Encrypt and Shuffle and EABEHP TransformCipherText algorithms perform exponential operations for each attribute, which significantly increases the algorithm execution time as the number of system attributes increases. On the other hand, the EABEHP Extract and ProxyDecrypt1 algorithm add multiplication operations as the number of system attributes increases, so the execution time of the algorithm slightly increases. From Table III, we can also see that the Encrypt and Shuffle algorithm occupies the majority of the execution time of the proposed EABEHP.

\vspace{-3.05mm}
	\subsection{Security Protocol Evaluation}\label{sec:protocolevaluation}
	\vspace{-0.5mm}
	
		\begin{figure}[t!]
		\begin{center}   
			{ 
				\includegraphics[scale=0.17]{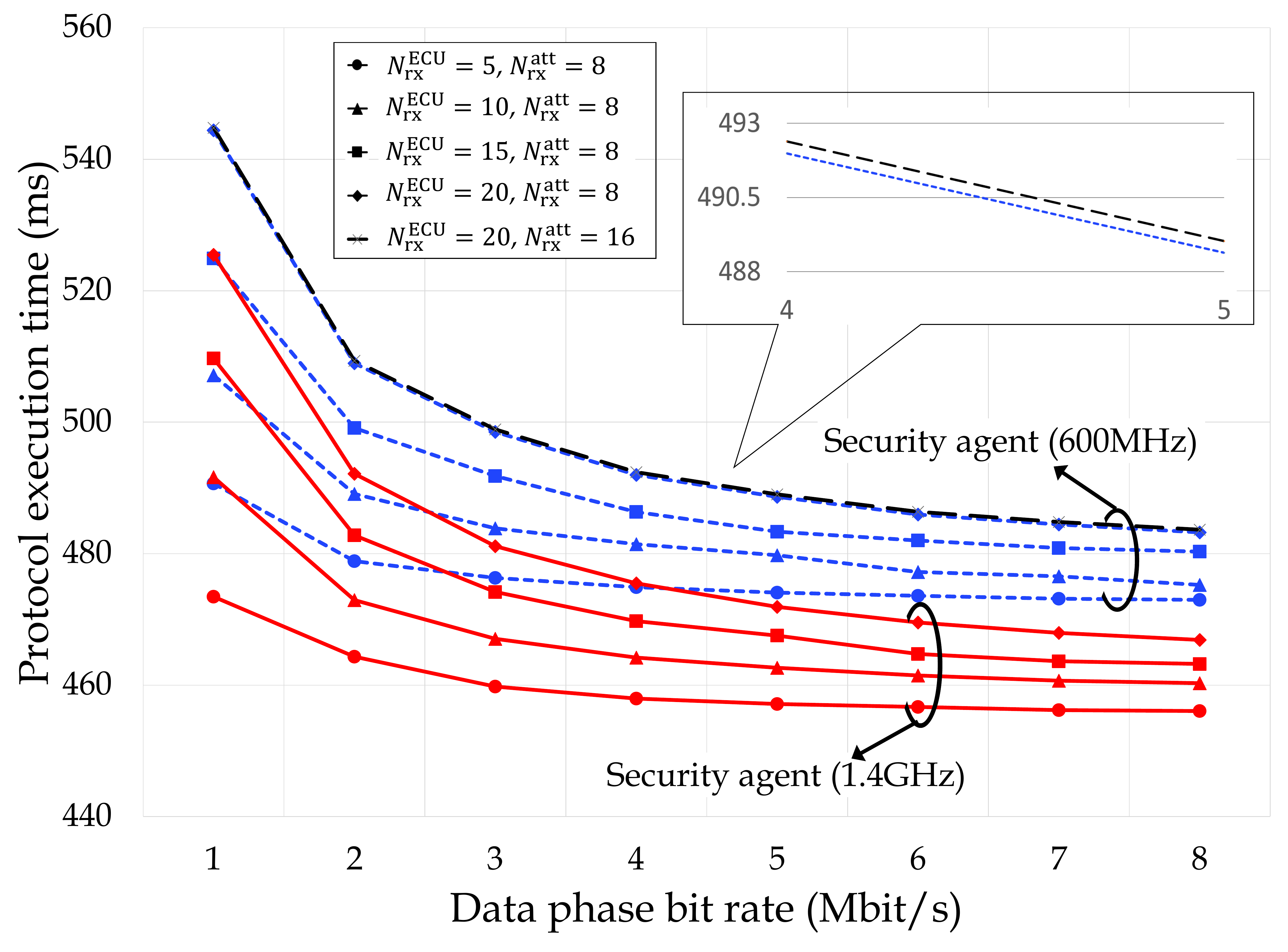}
			}
		\end{center}
		\vspace{-6mm}
		\caption{
			Execution time of the proposed protocol as a function of the data phase bit rate for different numbers of receiver-ECUs, $N_{\text{rx}}^{\text{ECU}}$, and receiver attributes, $N_{\text{rx}}^{\text{att}}$. Here, the number of system attributes is 16.
		}
		\label{fig:bitrate16}
		\vspace{-2.5mm}
	\end{figure}
	\begin{figure}[t!]
		\begin{center}   
			{ 
				\includegraphics[scale=0.17]{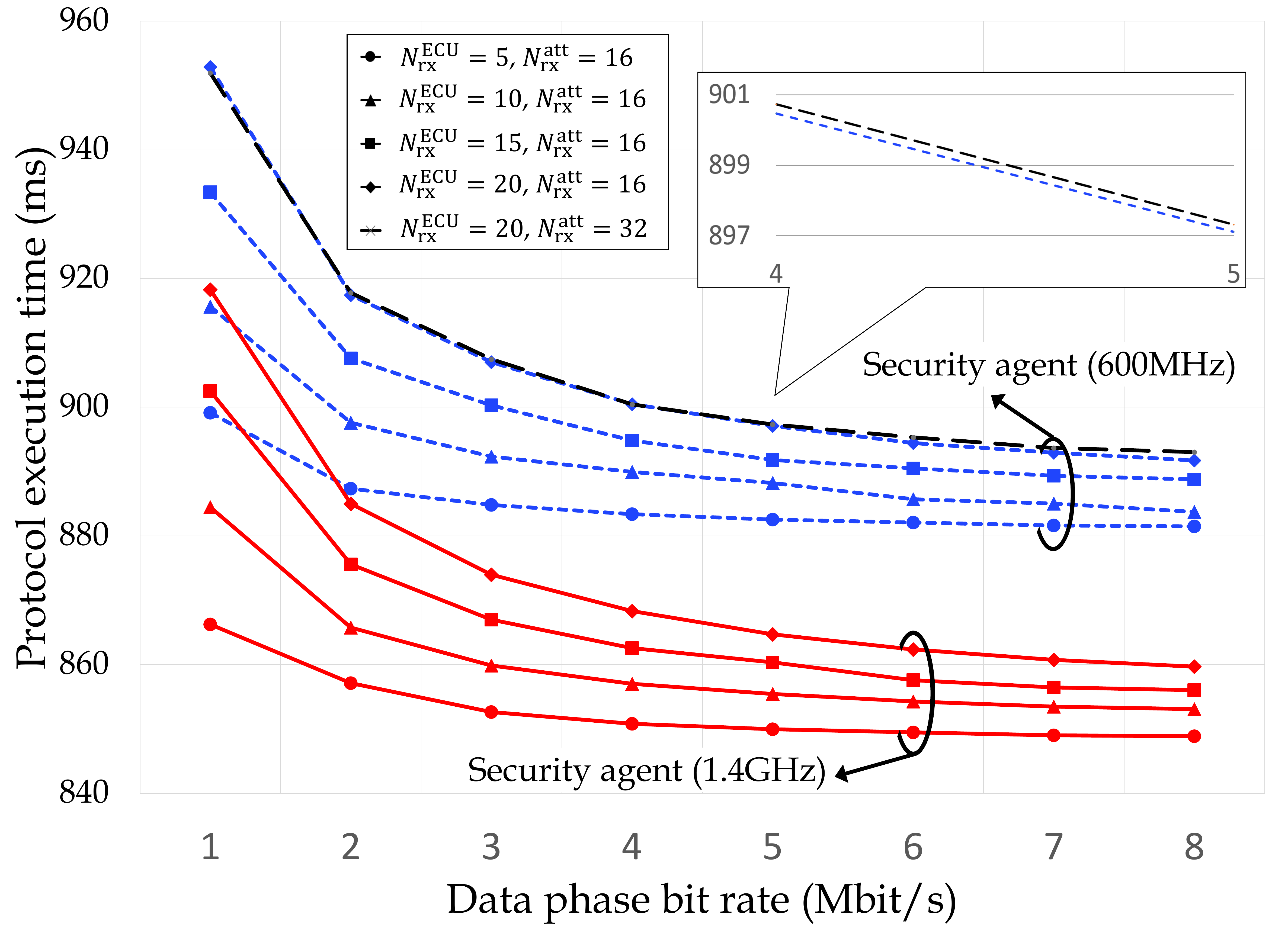}
			}
		\end{center}
		\vspace{-6mm}
		\caption{
			Execution time of the proposed protocol as a function of the data phase bit rate for different numbers of receiver-ECUs, $N_{\text{rx}}^{\text{ECU}}$, and receiver attributes, $N_{\text{rx}}^{\text{att}}$. Here, the number of system attributes is 32.
		}
		\label{fig:bitrate32}
		\vspace{-6mm}
	\end{figure} 	
	This subsection measures the execution time of the security protocol, based on the cryptographic algorithm evaluation. Using the CANoe v12.0 by Vector Co, we implement an evaluation environment similar to the real CAN-FD. The execution time of the proposed protocol is measured by considering the communication delay as well as the execution time of the cryptographic algorithms in Tables III and IV at the CANoe virtual ECU node. Note that, this work also achieves several additional features, such as attribute-based access control, and privacy-preserving for corrupted devices in addition to the security features achieved in existing in-vehicle security works. Furthermore, since this work proposes the novel edge computing-based security protocol that achieves a higher level of security and has reasonable latency in-vehicle systems, performance comparisons with other works are not included in the paper. Instead, we present the performance evaluation in various aspects to show that the proposed security protocol is practical.

	As shown in previous page, Figs. 5 and 6 show the execution time of the attribute-based authenticated key exchange protocols for different data phase bit rates. We perform the evaluation with the fixed arbitration phase bit rate of 0.5Mbit/s and adjust the data phase bit rate from 1Mbit/s to 8Mbit/s. 
	The number of system attributes and receiver attributes is set to 16 and 8, respectively, in Fig. 5, and 32 and 16, respectively, in Fig. 6. The measurement results when the SA clock speed is 1.4GHz and 600MHz are represented by the solid and dotted lines, respectively.
	
	From Figs. 5 and 6, we can see that the protocol execution time decreases as the data phase bit rate increases since the communication delay in CAN becomes smaller. By comparing the results with $N_{\text{rx}}^{\text{att}}=16$ and $N_{\text{rx}}^{\text{att}}=8$ in Fig. 5
	and those with $N_{\text{rx}}^{\text{att}}=32$ and $N_{\text{rx}}^{\text{att}}=16$ in Fig. 6, we can see that the number of the receiver attributes has little impact on the protocol execution time,	while the number of system attributes affects significantly on the protocol execution time.
	However, note that even with 32 system attributes, which is quite a large number to classify ECUs since there are many ECUs with overlapping roles, the execution time of the proposed protocol is less than 1 second.
	This means the proposed protocol can satisfy the practical requirements of in-vehicle networks. 
	
	Figure 7 shows the protocol execution time according to the number of system attributes for different numbers of system attributes, $N_{\text{sys}}^{\text{att}}$. The number of receiver attributes is set to be the same for all receivers, where the data phase bit rate is fixed at 4Mbit/s, the number of receiver-ECUs is 10, and the clock speed of the SA is 1.4GHz. We can see that the protocol execution time increases significantly as the number of system attributes increases while as mentioned above the number of receiver attributes has little effect on the protocol execution time. This is because the EABEHP Extract and ProxyDecrypt1, affected by the number of receiver attributes, are performed by a high-performance device, i.e., SA, and with relatively simple operations compared to other EABEHP algorithms. On the other hand, the EABEHP Encrypt algorithm, affected by the number of system attributes, is performed by the low-performance device, i.e., ECU, and with the complex operations.
	\begin{figure}[t!]
		\begin{center}   
			{ 
				\includegraphics[scale=0.42]{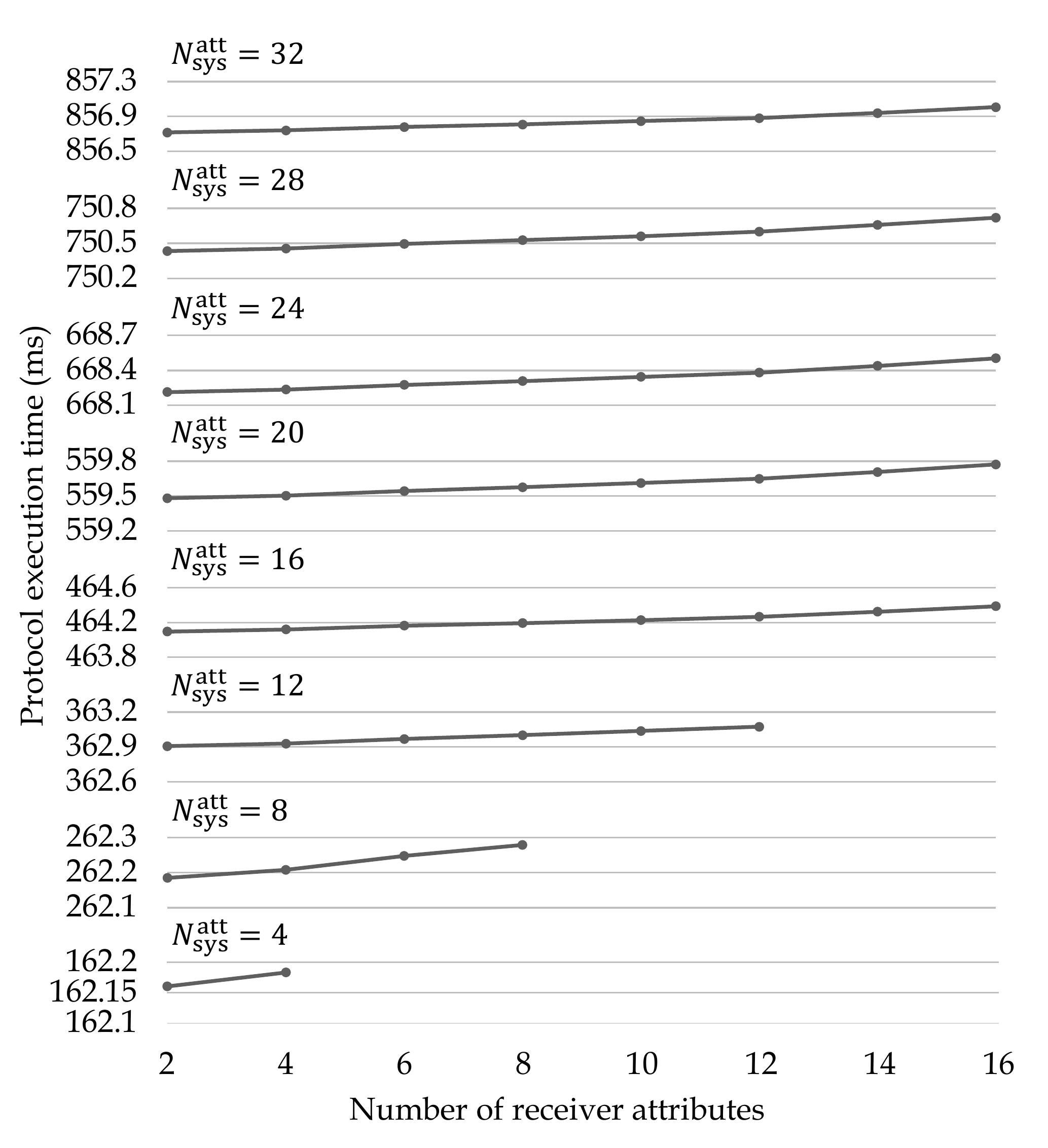}
			}
		\end{center}
	\vspace{-5mm}
		\caption{
			Execution time of the proposed protocol as a function of the number of receiver attributes, $N_{\text{rx}}^{\text{att}}$ for different numbers of system attribute, $N_{\text{sys}}^{\text{att}}$. 
			Here, the number of receiver-ECUs, $N_{\text{rx}}^{\text{ECU}}$ is 10. 
		}
		\label{fig:attribute}
		\vspace{-4.5mm}
	\end{figure}
	\begin{figure}[t!]
		\begin{center}   
			{ 
				\includegraphics[scale=0.27]{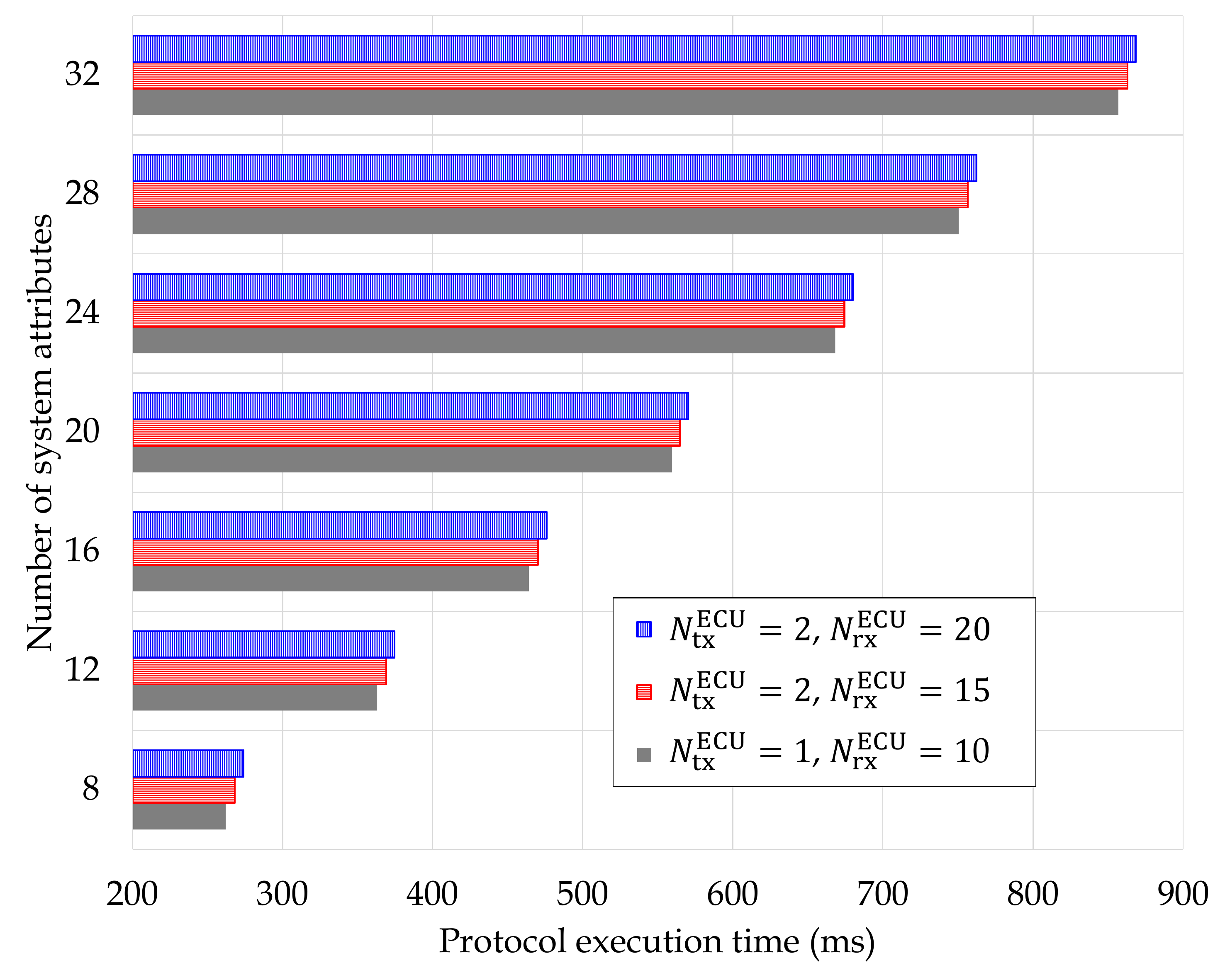}
			}
		\end{center}
	\vspace{-6mm}
		\caption{
			Execution time of the proposed protocol as a function of numbers of system attribute, $N_{\text{sys}}^{\text{att}}$, for different numbers of sender-ECUs, $N_{\text{tx}}^{\text{ECU}}$. 
			Here, the number of receiver-ECUs, $N_{\text{rx}}^{\text{ECU}}$ allocated to each sender-ECU is 10.
		}
		\label{fig:senderattribute}
		\vspace{-5.5mm}
	\end{figure}

	Figure 8 shows the protocol execution time according to the number of system attributes for different numbers of senders and receivers, denoted by $N_{\text{tx}}^{\text{ECU}}$ and $N_{\text{rx}}^{\text{ECU}}$, respectively. Here, we assume that 10 receivers access the message from one sender. The different number of senders and receivers are used : 1 sender and 10 receivers, 2 senders and 15 receivers (5 receivers received messages from both senders), and 2 senders and 20 receivers (all receivers received messages only from one sender).
	We can see that the number of system attributes have a little impact on the gap in protocol execution time for the above three cases. This is because, when the priority of the message is well-established, the cryptographic algorithm execution time, associated with the number of system attributes, is generally much longer than the communication delay. Here, the increase in communication delay due to the increase in the number of senders who simultaneously transmit the message is negligible.
		\begin{figure}[t!]
		\begin{center}   
			{ 
				\includegraphics[scale=0.28]{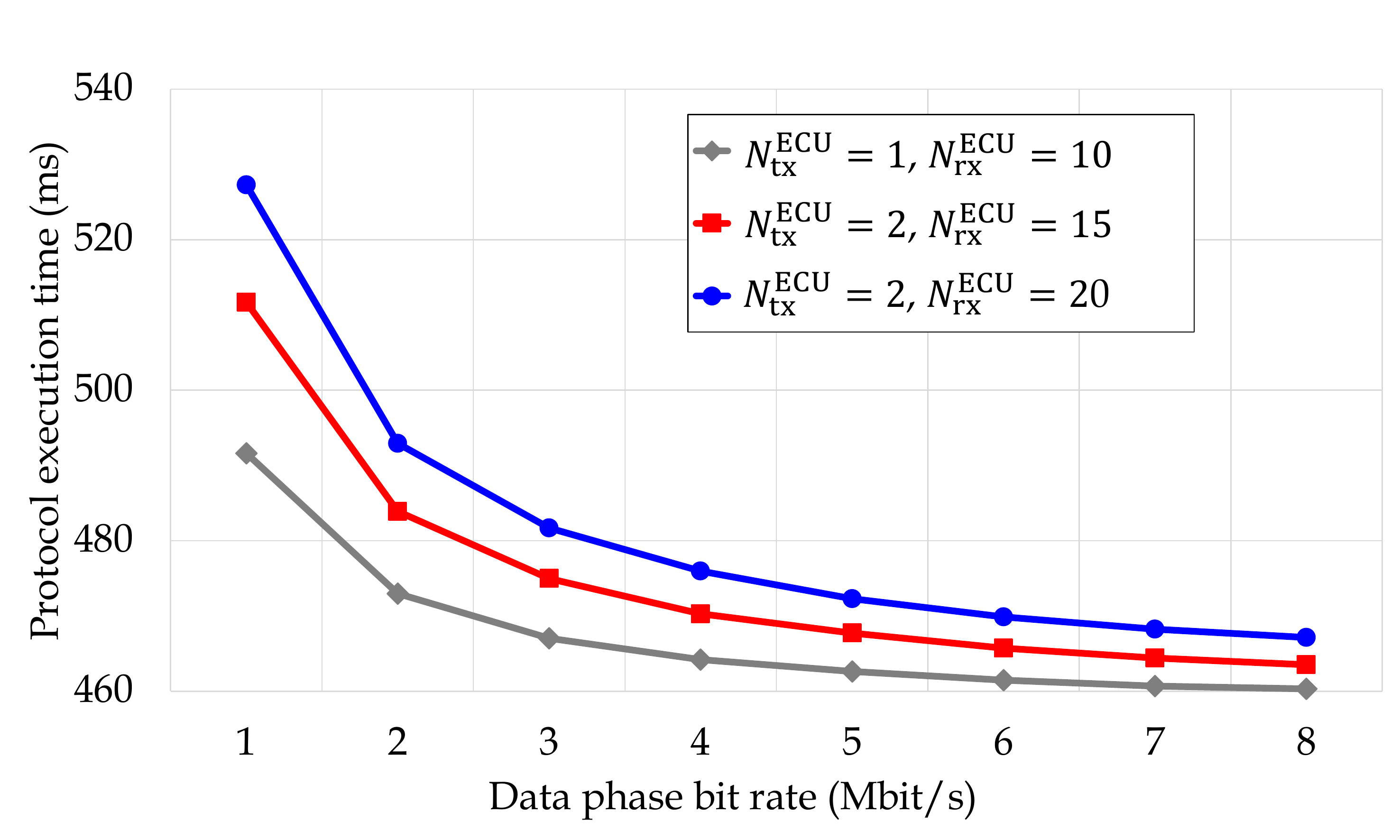}
			}
		\end{center}
		\vspace{-6mm}
		\caption{
			Execution time of the proposed protocol as a function of the data phase bit rate, for different numbers of sender-ECUs, $N_{\text{tx}}^{\text{ECU}}$. 
			Here, the number of receiver-ECUs, $N_{\text{rx}}^{\text{ECU}}$ allocated to each sender-ECU is 10. 
		}
		\label{fig:senderbitrate}
		\vspace{-6mm}
	\end{figure}
	Figure 9 shows the protocol execution time versus the data phase bit rate for the above three cases. We can see that the protocol execution time difference among three cases becomes smaller as the data phase bit rate increases. This is because the number of communications is different in three cases, and the communication delay is inversely proportional to the data phase bit rate. Hence, the larger the data phase bit rate gives the less the protocol execution time difference.
	Through the results of Figs. 8 and 9, we can see that if the data phase bit rate is high enough, the time taken to execute the proposed attribute-based key exchange process on all ECUs will not be significantly affected by the in-vehicle network size, i.e., the numbers of sender-ECUs and receiver-ECUs. Therefore, the proposed protocol is expected to be executed in a reasonable time even for different in-vehicle network sizes, which shows the feasibility and the practicality of the proposed protocol.

	\vspace{-3mm}
	\section{CONCLUSION}\label{sec:conclusion}
	This work proposes an edge computing-based in-vehicle security protocol with the attribute-based access control that privacy for policy and credentials.
	The security of this protocol has been proved through security analysis to be limited to the security of pseudorandom function, pseudorandom permutation, and C-IND-CPA-RUCA and P-IND-CPA-UCA EABEHP. 
	Specifically, the performance analysis of the proposed protocol shows the effect of protocol execution time according to the data phase bit rate, the number of system attributes, the number of receiver attributes, and the number of sender and receiver-ECUs. This shows that a high-security level can be satisfied in an appropriate latency in an in-vehicle communication environment having a resource-poor ECU.
	Hence, this work has demonstrated to support efficient secure communication with fine-grained access control for in-vehicle networks.
	\vspace{-3.5mm}
	\vspace{-1.5mm}
	

\end{document}